\documentclass[journal]{IEEEtran}

\ifCLASSINFOpdf
   \usepackage[pdftex]{graphicx}
\else
  \usepackage[dvips]{graphicx}
\fi

\usepackage[cmex10]{amsmath}
\usepackage{amssymb}
\usepackage{caption}
\usepackage{amsthm}
\usepackage{tikz}
\usepackage{pgfplots}
\pgfplotsset{compat=newest}
\pgfplotsset{legend style={rounded corners=2pt,nodes=right}}
\usetikzlibrary{arrows,decorations,shapes}
\usepackage{color,cite}
\DeclareMathAlphabet{\mathbit}{OML}{cmr}{bx}{it}

\DeclareMathOperator{\Q}{Q}
\DeclareMathOperator{\E}{E}

\newcommand\Sign{\operatorname{sign}}

\DeclareMathOperator{\T}{\operatorname{T}}

\DeclareMathOperator{\fieldR}{\mathbb{R}}
\DeclareMathOperator{\fieldB}{\mathbb{B}}

\newcommand{\ve}[1]{\boldsymbol{#1}}

\newcommand{\exdi}[2]{\E_{#1} \left[#2\right]}
\newcommand{\exdiinv}[2]{\E_{#1}^{-1} \left[#2\right]}

\newcommand{\expb}[1]{\exp{\left(#1\right)}}

\newcommand{\sign}[1]{\Sign{\left(#1\right)}}

\newcommand{\qfunc}[1]{\Q \left(#1\right)}
\newcommand{\qfuncptwo}[1]{\Q^2 \left(#1\right)}
\DeclareMathAlphabet\mathbfcal{OMS}{cmsy}{b}{n}
\newtheorem{theorem}{Theorem}

\begin{document}

\title{Performance Analysis for Channel Estimation with 1-bit ADC and Unknown Quantization Threshold}

\author{Manuel S. Stein, Shahar Bar, Josef A. Nossek, and Joseph Tabrikian
\thanks{The work leading to this publication was supported by the German Academic Exchange Service (DAAD) with funds from the German Federal Ministry of Education and Research (BMBF) and the People Program (Marie Sk{\l}odowska-Curie Actions) of the European Union's Seventh Framework Program (FP7) under REA grant agreement 605728 (P.R.I.M.E. - Postdoctoral Researchers International Mobility Experience). The work leading to this publication was also supported by the Israel Science Foundation (Grant 1160/15) and by the Yaakov ben Yitzhak Hacohen Scholarship.}
\thanks{
M. S. Stein is with Mathematics Department, Vrije Universiteit Brussel, Belgium, and also with the Chair for Stochastics, Universit\"at Bayreuth, Germany (e-mails: manuel.stein@uni-bayreuth.de). S. Bar and J. Tabrikian are with the Department of Electrical and Computer Engineering,  Ben-Gurion University of the Negev, Beer-Sheva, Israel (e-mail: shahba@post.bgu.ac.il, joseph@bgu.ac.il). J. A. Nossek is with the Department of Teleinformatics Engineering, Universidade Federal do Cear\'{a}, Brasil, and also with the Department of Electrical and Computer Engineering, Technische Universit\"at M\"unchen, Germany (e-mail: josef.a.nossek@tum.de). }
}
\maketitle
\begin{abstract}
In this work, the problem of signal parameter estimation from measurements acquired by a low-complexity analog-to-digital converter (ADC) with $1$-bit output resolution and an unknown quantization threshold is considered. Single-comparator ADCs are energy-efficient and can be operated at ultra-high sampling rates. For analysis of such systems, a fixed and known quantization threshold is usually assumed. In the symmetric case, i.e., zero hard-limiting offset, it is known that in the low signal-to-noise ratio (SNR) regime the signal processing performance degrades moderately by ${2}/{\pi}$ ($-1.96$ dB) when comparing to an ideal $\infty$-bit converter. Due to hardware imperfections, low-complexity $1$-bit ADCs will in practice exhibit an unknown threshold different from zero. Therefore, we study the accuracy which can be obtained with receive data processed by a hard-limiter with unknown quantization level by using asymptotically optimal channel estimation algorithms. To characterize the estimation performance of these nonlinear algorithms, we employ analytic error expressions for different setups while modeling the offset as a nuisance parameter. In the low SNR regime, we establish the necessary condition for a vanishing loss due to missing offset knowledge at the receiver. As an application, we consider the estimation of single-input single-output wireless channels with inter-symbol interference and validate our analysis by comparing the analytic and experimental performance of the studied estimation algorithms. Finally, we comment on the extension to multiple-input multiple-output channel models.
\end{abstract}
\begin{IEEEkeywords}
$1$-bit ADC, Cram\'er-Rao bounds, channel estimation, hard-limiting loss, intersymbol interference, nuisance parameter, quantization threshold, wireless communication
\end{IEEEkeywords}
\IEEEpeerreviewmaketitle
\section{Introduction}
\IEEEPARstart{T}{HE} design of signal processing systems is governed by two conflicting objectives. On the one hand, an architecture which allows obtaining high operational performance and small latency is desired. On the contrary, the processing device should exhibit low complexity concerning its power consumption, production cost, and circuit size. In connection with the latter aspect, it has been identified that analog-to-digital (A/D) conversion forms a bottleneck \cite{Walden99,Verhelst15}. In this step, continuous waveforms (analog signal domain), acquired at the receive sensors, are transformed into a representation which is discrete in time and amplitude (digital signal domain). The resulting data can then be processed by sophisticated algorithms which are executed by dedicated hardware or by a general-purpose chip. As the complexity of the A/D conversion can grow exponentially $\mathcal{O}(2^b)$ with the number of bits $b$ which are used for the representation of the amplitude information, the A/D resolution restricts the receive bandwidth and significantly affects the overall energy consumption. This is, in particular, an issue if sampling rates above $100$ MHz have to be realized\cite{MurmannSurvey}. Thus, an interesting option, especially for low-cost wide-band receivers in the Internet of things (IoT) \cite{Landau17} or high-performance base-stations with massive antenna arrays in mobile communication \cite{Choi16}, is to switch to coarse A/D resolution. Precise physical modeling, adapted system design and digital signal processing with powerful nonlinear algorithms then allows coping with the envisioned accuracy and throughput requirements. 

\subsection{Low-Complexity $1$-bit A/D Conversion}
A radical approach is to use a single comparator, which only forwards the sign to the digital domain and discards all the information about the analog signal amplitude. This approach results in a cheap, small, and fast A/D converter (ADC) with low energy consumption. Additionally, an automatic gain control (AGC) circuit \cite{Khoury98} is not required. Due to these attractive properties, a vivid discussion on $1$-bit quantization has emerged in the field of modern signal processing \cite{Jacovitti94,Lok98,Madsen00,BarShalom02,Boufounos08,Mez10,Dabeer10}, while \cite{Bennett48,Vleck66,Curry70} are classical references for this topic. Furthermore, communication over channels with $1$-bit quantizer is considered in recent works \cite{Dabeer06,Ivra07,Mez12,Mo15,Jacobsson15,Bender16}.

Despite its low complexity, $1$-bit A/D conversion introduces nonlinearity into the system model, which is associated with a performance loss. When the system operates in the low signal-to-noise ratio (SNR) regime, the loss is moderate with ${2}/{\pi}$ or $-1.96$ dB \cite{Vleck66}. Further, the simplicity of the radio front-end allows exploiting other design options which are crucial for system performance. For example, faster sampling rates \cite{Gilbert93,Shamai94,Koch10,Krone12,Zhang12,Landau14,SteinICASSP17} or a higher number of receive sensors \cite{SteinPLANS14,SteinWSA16} allow reducing the hard-limiting loss. Also, the analog pre-filter \cite{SteinICASSP13} or the demodulator \cite{SteinWCL15} can be adjusted to diminish the nonlinear loss with coarse resolution ADCs. Taking into account side-information about the temporal evolution of the channel parameters is also an effective approach to obtain high accuracy with $1$-bit A/D conversion \cite{SteinTSP15}. 

Another line of work deals with the optimization of the 1-bit ADC by modification of the quantization level. In \cite{Balkan10} it is shown that for pilot-based channel estimation a deterministic time-varying hard-limiting threshold yields a higher Fisher information than a random offset and therefore enables to minimize the estimation error. The discussion in \cite{Koch13} aims at higher communication rates and studies maximization of Shannon information with asymmetric $1$-bit quantization at the receiver. In contrast, the works \cite{Dabeer06_2,Dabeer08,Zeitler12} consider dithering, i.e., controlled randomization of the quantization level. 

Channel estimation with coarsely quantized data is considered in \cite{Jacobsson15,Li17,Molen17}, where linear estimation techniques are used after symmetric hard-limiting, while \cite{Ivra07} studies nonlinear reconstruction of the unquantized receive signal with a subsequent linear channel estimator. Under symmetric quantization with arbitrary bits \cite{Mezghani12} proposes a message-passing algorithm, while for symmetric hard-limiting, \cite{Choi16} proposes nonlinear likelihood-based channel estimation. The work \cite{Wu09} discusses distributed channel estimation with iterative adaptation of the quantization threshold.

\subsection{Motivation and Contribution}
In practice, change of the quantization level during runtime requires to feedback analog control information to an offset voltage source. As the control signal has to be determined in the digital domain, a digital-to-analog converter (DAC) with high output resolution is required for the accurate adjustment of the quantization level. Since also the complexity of DACs grows significant with the number of input bits, such an approach stands in contradiction to the goal of minimizing the radio front-end complexity by $1$-bit A/D technology. Therefore, for low-cost receivers, the design of the $1$-bit A/D conversion will be such that the offset is close to a predetermined constant value. Hardware imperfections, variations in the production process, and external effects, as discussed in \cite{Kurosawa01}, lead to the situation that the quantization offset is in general unknown. Therefore, calibration or a method which determines and compensates the offset during runtime is required.

In this paper, the problem of channel parameter estimation, subject to measurements obtained with a $1$-bit ADC under an unknown quantization threshold is studied. To the best of our knowledge, this particular problem has not been addressed in previous works. To provide a thorough discussion, we focus on single-input single-output (SISO) channels and take two different modeling perspectives. First, with the mindset of frequentists \cite{Chaumette,Ren1,Menni}, we assume that the parameters of interest and the quantization threshold are deterministic unknown variables. Then, we consider a hybrid model \cite{HCRLB,Reuven,Noam,Messer,Ren}, where the channel parameters are random and distributed according to a known probability distribution function, while the quantization offset is modeled as a deterministic unknown nuisance parameter. The hybrid approach is motivated by the fact that in various cases prior information about the channel is available at the receiver. This information can be incorporated into the inference process to improve the estimation accuracy. 

For both situations, we use asymptotically optimal estimation algorithms and study their performance by asymptotic expressions. In particular, we investigate the performance gaps between an ideal receiver with infinite ADC resolution, a low-complexity receive system employing a $1$-bit ADC with a known threshold, and a receiver where the quantization threshold is unknown. In the low SNR regime, we establish the result that missing offset knowledge does not degrade the estimation accuracy. For a wireless SISO channel with inter-symbol interference (ISI), we verify the results by Monte-Carlo simulations. These experimental findings show that the conducted analysis accurately captures the performance trends of signal processing applications with $1$-bit ADC and unknown quantization offset. In the end, we briefly comment on the generalization of the analysis to multiple-input multiple-output (MIMO) channels. The presented results are an extension of our conference contribution \cite{SteinBar_ICASSP2016}, which was confined to an observation model with a scalar channel parameter.

\subsection{Outline}
This paper is organized as follows. In Section \ref{sec:Basics} we define the receive models without and with $1$-bit A/D conversion. Section \ref{sec:Bound} discusses a deterministic and a hybrid modeling framework for the channel estimation task, outlines the asymptotically optimal estimation procedures and investigates their performance by analytic expressions. The estimation accuracy for operation in the low SNR regime is studied in Section \ref{sec:lowSNR}, whereas in Section \ref{sec:sim} we demonstrate the results for the application of wireless channel estimation with intersymbol interference. Additionally, we validate our theoretic findings by Monte-Carlo simulations of practical signal processing algorithms. The conclusions appear in Section \ref{sec:conc}.
\section{System Model}\label{sec:Basics}
We consider two different system models. The first receive system features an ADC with $b$-bits output resolution, where $b$ is sufficiently high such that the effect of amplitude quantization can be neglected. For simplicity, in the following we will refer to this setup as an ideal receiver with $\infty$-bit ADC. The second system is a low-complexity receiver with $1$-bit ADC resolution, where after the A/D conversion only binary information about the analog receive signal amplitude is available for further digital signal processing.
\subsection{Ideal Receive System}
The receive signal of the $\infty$-bit system at time instant $n$ with $n=1,\ldots,N$ is modeled by the random variable ${y}_n\in\fieldR$, 
\begin{align}\label{system:model:ideal:generic}
{y}_n \sim p_{{y}_n}({y}_n|\ve{\theta}),
\end{align}
following a Gaussian conditional probability density function
\begin{align}\label{eq:condpdf:ideal}
p_{{y}_n}({y}_n | \ve{\theta})&=(2\pi)^{-\frac{1}{2}}\expb{-\frac{1}{2} \big({y}_n- {s}_n( \ve{\theta})\big)^2},
\end{align}
where ${s}_n( \ve{\theta})\in\fieldR$ is a pilot sequence of deterministic structure. The signal ${s}_n( \ve{\theta})$ is modulated by the channel parameters, summarized in the vector $\ve{\theta}\in\ve{\Theta}\subset\fieldR^K$. Further, ${s}_n( \ve{\theta})$ is continuously differentiable with respect to $\ve{\theta}$. Since one can always normalize the receive signal by its standard deviation, without loss of generality, the variance of \eqref{eq:condpdf:ideal} is assumed to be $1$. The data model \eqref{system:model:ideal:generic} can be extended to complex-valued receiver models by considering two independent real-valued random variables. As this has no impact onto the presented results, for the sake of simplicity, we focus on the real-valued case. To explicitly focus on the effect of threshold estimation, during the discussion we assume white additive Gaussian noise like commonly done in the signal processing and wireless communication literature. 

\subsection{Low-Complexity Receive System}
The receiver with $1$-bit A/D conversion can be modeled
\begin{align}\label{system:model:sign}
{z}_n = \sign{{y}_n-{\alpha}},
\end{align}
where $\sign{x}$ is the signum function defined as
\begin{align}
\sign{x}\triangleq
\begin{cases}
+1& \text{if } x \geq 0\\
-1& \text{if } x < 0
\end{cases},
\end{align}
${\alpha}\in\fieldR$ forms an unknown deterministic quantization threshold, and $\triangleq$ denotes equality by definition. 

Concerning the quantization model \eqref{system:model:sign}, note that we refer to an A/D conversion without a feedback loop. This distinguishes the topic of low-complexity $1$-bit ADCs from the sigma-delta modulation approach, in which a single comparator with feedback is operated in a highly oversampled mode to perform the A/D conversion \cite{Boser88,Aziz96}.

The quantized observation model \eqref{system:model:sign} is characterized by a binary random variable $z_n\in\fieldB\triangleq\{-1,1\}$,
\begin{align}\label{system:model:1bit:generic}
{z}_n \sim p_{{z}_n}({z}_n|\ve{\psi}),
\end{align}
following the conditional probability mass function
\begin{align}\label{eq:condpdf:1bit}
p_{{z}_n}({z}_n|\ve{\psi}) &= \qfunc{ z_{n}\big(\alpha - s_{n}(\ve{\theta})\big)},
\end{align}
where 
\begin{align}\label{def:qfunction}
\qfunc{x}\triangleq\frac{1}{\sqrt{2\pi}} \int_{x}^{\infty} \exp{\left(-\frac{u^2}{2}\right)} {\rm d} u
\end{align}
denotes the Q-function. The probability mass function \eqref{eq:condpdf:1bit} is parametrized by the unknown vector parameter 
\begin{align}
\boldsymbol{\psi}\triangleq\begin{bmatrix} \ve{\theta}^{\T} & \alpha\end{bmatrix}^{\T}\in\ve{\Psi}\triangleq\ve{\Theta}\times\fieldR, 
\end{align}
of which $\ve{\theta}$ serves as the parameter of interest, while the offset $\alpha$ forms a nuisance parameter. Note, that we do not distinguish between probability mass \eqref{system:model:1bit:generic} and probability density functions \eqref{system:model:ideal:generic}. The respective case is always clear from the context.

\subsection{Signal Processing Task}
The signal processing task of the receivers is to calculate the estimates $\ve{\hat{\theta}}_{\ve{y}}(\ve{y})$ and $\ve{\hat{\theta}}_{\ve{z}}(\ve{z})$ by using the $N$ receive samples
\begin{align}
\ve{y}=\begin{bmatrix} {y}_1 &{y}_2 &\ldots {y}_N\end{bmatrix}^{\T},
\end{align}
or
\begin{align}
\ve{z}=\begin{bmatrix} {z}_1 &{z}_2 &\ldots {z}_N\end{bmatrix}^{\T}
\end{align}
and the available knowledge about the models \eqref{eq:condpdf:ideal} and \eqref{eq:condpdf:1bit}.
\section{Theory - Performance Analysis}\label{sec:Bound}
To characterize the performance gap between both systems, we discuss two different settings. For each of them, we review the optimum estimation algorithm for the asymptotic regime and establish the achievable estimation performances by analytical error bounds or asymptotic error expressions.
\subsection{Deterministic Modeling Approach}
First, we study the case where the channel parameters $\ve{\theta}$ and the threshold ${\alpha}$ are both deterministic but unknown. Under these assumptions, we evaluate the performance of the estimators $\ve{\hat{\theta}}_{\ve{y}}(\ve{y})$ and $\ve{\hat{\theta}}_{\ve{z}}(\ve{z})$ under the mean squared error (MSE) criterion,
\begin{align}\label{definition:mse:unquantized:receiver}
\ve{\text{MSE}}_{\ve{y}}(\ve{\theta} )&\triangleq\exdi{\ve{y}|\ve{\theta}}{\big(\ve{\hat{\theta}}_{\ve{y}}(\ve{y})-\ve{\theta}\big) \big(\ve{\hat{\theta}}_{\ve{y}}(\ve{y})-\ve{\theta}\big)^{\T}},\\
\label{definition:mse:quantized:receiver}
\ve{\text{MSE}}_{\ve{z}}(\ve{\psi} )&\triangleq\exdi{\ve{z}|\ve{\psi}}{\big(\ve{\hat{\theta}}_{\ve{z}}(\ve{z})-\ve{\theta}\big) \big(\ve{\hat{\theta}}_{\ve{z}}(\ve{z})-\ve{\theta}\big)^{\T}},
\end{align}
where the MSE of the $1$-bit receiver \eqref{definition:mse:quantized:receiver} is a function of the parameters $\ve{\theta}$ and the offset ${\alpha}$.
\subsubsection{Estimation Procedure}
Concerning the performance characterizations \eqref{definition:mse:unquantized:receiver} and \eqref{definition:mse:quantized:receiver}, the asymptotically optimum unbiased estimator with both receivers is the maximum-likelihood estimator (MLE) \cite{Kay93}, given by
\begin{align}\label{def:mle:ideal}
\ve{\hat{\theta}}_{\ve{y}}(\ve{y})&\triangleq\arg \max_{\ve{\theta}\in \ve{\Theta}} p_{\ve{y}}(\ve{y}|\ve{\theta})\notag\\
&=\arg \max_{\ve{\theta}\in \ve{\Theta}} \sum_{n=1}^{N} \ln p_{{y}_n}({y}_n|\ve{\theta})
\end{align}
for the unquantized case and
\begin{align}\label{definition:mle:quantized}
\begin{bmatrix}\ve{\hat{\theta}}_{\ve{z}}^{\T}(\ve{z}) &\hat{\alpha}(\ve{z})\end{bmatrix}^{\T}&\triangleq\arg \max_{\ve{\psi}\in \ve{\Psi}} p_{\ve{z}}(\ve{z}|\ve{\theta},{\alpha})\notag\\
&=\arg \max_{\ve{\psi} \in \ve{\Psi}} \sum_{n=1}^{N} \ln p_{{z}_n}({z}_n|\ve{\theta},{\alpha})
\end{align}
for the low-complexity $1$-bit ADC receiver. Note that for the $1$-bit system, the estimation of the channel parameters $\ve{\theta}$ and the hard-limiting offset ${\alpha}$ has to be performed jointly.

Under some mild regularity conditions (see \cite{BarShalom,Bhat,Silvey}), in the asymptotic regime, the MSE of the $\infty$-bit receiver in \eqref{definition:mse:unquantized:receiver} implementing the MLE, is given by the Cram\'{e}r-Rao lower bound (CRLB) \cite{Rao45,Cram46}
\begin{align}
\label{eq:CRLB}
\ve{\text{MSE}}_{\ve{y}}(\ve{\theta} )\overset{a}{=} \ve{F}^{-1}(\ve{\theta}),
\end{align}
where, under the notational convention
\begin{align}
\left[ \frac{\partial \ve{g}(\ve{x})}{\partial \ve{x}} \right]_{ij} \triangleq \frac{\partial g_i(\ve{x})}{\partial x_j},
\end{align}
the Fisher information matrix (FIM) is defined \cite{Kay93}
\begin{align}
\ve{F}(\ve{\theta})&\triangleq\exdi{\ve{y}|\ve{\theta}}{ \bigg(\frac{\partial \ln p_{\ve{y}}(\ve{y}|\ve{\theta}) }{\partial \ve{\theta}} \bigg)^{\T} \frac{\partial \ln p_{\ve{y}}(\ve{y}|\ve{\theta}) }{\partial \ve{\theta}} }
\label{info:fishery}
\end{align}
and $\overset{a}{=}$ is used to denote asymptotic equality, i.e., equality after taking the number of samples $N$ to infinity. Note that due to the statistical independence of the samples in \eqref{eq:condpdf:ideal}, we have
\begin{align}\label{eq:FIM:independence}
\ve{F}(\ve{\theta})&=\sum_{n=1}^{N}  \ve{F}_n(\ve{\theta}),
\end{align}
where
\begin{align}
\ve{F}_n(\ve{\theta})&\triangleq \exdi{{y}_n|\ve{\theta}}{ \bigg(\frac{\partial \ln p_{{y}_n}({y}_n|\ve{\theta}) }{\partial \ve{\theta}} \bigg)^{\T} \frac{\partial \ln p_{{y}_n}({y}_n|\ve{\theta}) }{\partial \ve{\theta}} }\notag\\
&=  \left( \frac{\partial {s}_n(\ve{\theta})}{\partial \ve{\theta}}\right)^{\T}   \frac{\partial {s}_n(\ve{\theta})}{\partial \ve{\theta}}\notag\\
&=\ve{f}^{\T} _n(\ve{\theta})\ve{f}_n(\ve{\theta})
\label{info:fishery:sample}
\end{align}
with the notational convention
\begin{align}\label{eq:smallFIM:sample}
\ve{f}_n(\ve{\theta})&\triangleq\left( \frac{\partial {s}_n(\ve{\theta})}{\partial \ve{\theta}}\right)^{\T}.
\end{align}

For the $1$-bit receiver \eqref{eq:condpdf:1bit}, the FIM exhibits a block structure
\begin{align}
\ve{J}(\ve{\psi})=\begin{bmatrix} \ve{J}_{\ve{\theta}\ve{\theta}}(\ve{\psi}) &\ve{J}_{\ve{\theta}\alpha}(\ve{\psi})\\
\ve{J}_{\alpha\ve{\theta}}(\ve{\psi}) &{J}_{\alpha\alpha}(\ve{\psi}) \end{bmatrix},\label{eq:FIM:quant}
\end{align}
such that the asymptotic MSE of the MLE estimator $\ve{\hat{\theta}}_{\ve{z}}(\ve{z})$ is equivalent to the CRLB
\begin{align}
\label{eq:CRLB:1bit}
&\ve{\text{MSE}}_{\ve{z}}(\ve{\psi} )\overset{a}{=}\bigg( \ve{J}_{\ve{\theta}\ve{\theta}}(\ve{\psi}) -  \frac{\ve{J}_{\ve{\theta}\alpha}(\ve{\psi})    \ve{J}_{\alpha\ve{\theta}}(\ve{\psi})}{{J}_{\alpha\alpha} (\ve{\psi})}   \bigg)^{-1},
\end{align}
where the expressions required in \eqref{eq:FIM:quant} are given by
\begin{align}\label{eq:Jtt}
\ve{J}_{\ve{\theta}\ve{\theta}}(\ve{\psi})&\triangleq\exdi{\ve{z}|\ve{\psi}}{ \bigg(\frac{\partial \ln p_{\ve{z}}(\ve{z}|\ve{\psi}) }{\partial \ve{\theta}} \bigg)^{\T} \frac{\partial \ln p_{\ve{z}}(\ve{z}|\ve{\psi}) }{\partial \ve{\theta}} },\\
\ve{J}_{\ve{\theta}\alpha}(\ve{\psi})
&\triangleq\exdi{\ve{z}|\ve{\psi}}{ \bigg(\frac{\partial \ln p_{\ve{z}}(\ve{z}|\ve{\psi}) }{\partial \ve{\theta}} \bigg)^{\T} \frac{\partial \ln p_{\ve{z}}(\ve{z}|\ve{\psi}) }{\partial {\alpha}} },\label{eq:Jat}\\
\ve{J}_{\alpha\ve{\theta}}(\ve{\psi})&\triangleq\ve{J}_{\ve{\theta}\alpha}^{\T}(\ve{\psi}),
\end{align}
and 
\begin{align}\label{eq:Jaa}
{J}_{\alpha\alpha}(\ve{\psi})&\triangleq\exdi{\ve{z}|\ve{\psi}}{\bigg(\frac{\partial \ln p_{\ve{z}}(\ve{z}|\ve{\psi}) }{\partial {\alpha}} \bigg)^2}.
\end{align}
Note that we use the letter $\ve{J}$ for the FIMs associated with the quantized receiver \eqref{system:model:sign} to clearly distinguish from the FIMs $\ve{F}$ associated with the ideal receiver \eqref{system:model:ideal:generic}. 

Using \eqref{eq:condpdf:1bit} and the derivative 
\begin{align}
\frac{\partial \qfunc{x}}{\partial x}= -\frac{1}{\sqrt{2\pi}} \exp{\left(-\frac{x^2}{2}\right)} 
\end{align}
of the Q-function \eqref{def:qfunction}, we obtain
\begin{align}\label{eq:condpdf:1bit:derivative}
\frac{\partial \ln p_{{z}_n}(z_{n}|\ve{\psi}) }{\partial \ve{\theta}} =   \frac{z_n \exp{ \big(-\frac{(\alpha- s_{n}(\ve{\theta}))^2}{2} \big)} }{\sqrt{2 \pi}\qfunc{z_n(\alpha- s_{n}(\ve{\theta}))} } \frac{\partial s_{n}(\ve{\theta})}{\partial \ve{\theta}}.
\end{align}
Using the expressions \eqref{eq:condpdf:1bit} and \eqref{eq:condpdf:1bit:derivative}, with
\begin{align}\label{definition:phin}
\phi_n(\ve{\psi})\triangleq\frac{ \expb{ -\big(\alpha- s_{n}(\ve{\theta})\big)^2 }  }{ {2\pi} \big( \qfunc{\alpha- s_{n}(\ve{\theta})} - \qfuncptwo{\alpha- s_{n}(\ve{\theta})} \big) },
\end{align}
we derive
\begin{align}\label{derivation:FIM:quantized}
&\exdi{{z}_n|\ve{\psi}}{ \bigg(\frac{\partial \ln p_{{z}_n}({z}_n|\ve{\psi}) }{\partial \ve{\theta}} \bigg)^{\T} \frac{\partial \ln p_{{z}_n}({z}_n|\ve{\psi}) }{\partial \ve{\theta}} }=\notag\\
&=\phi_n(\ve{\psi})\left( \frac{\partial s_{n}(\ve{\theta})}{\partial \ve{\theta}}\right)^{\T}  \frac{\partial s_{n}(\ve{\theta})}{\partial \ve{\theta}},
\end{align}
where the step-by-step calculation is given in Appendix \ref{app:derivation:FIM}. 

Therefore, we can write the first FIM block from \eqref{eq:Jtt} as
\begin{align}
\ve{J}_{\ve{\theta}\ve{\theta}}(\ve{\psi})&=
\sum_{n=1}^{N} \exdi{{z}_n|\ve{\psi}}{ \bigg(\frac{\partial \ln p_{{z}_n}({z}_n|\ve{\psi}) }{\partial \ve{\theta}} \bigg)^{\T} \frac{\partial \ln p_{{z}_n}({z}_n|\ve{\psi}) }{\partial \ve{\theta}} }\notag\\
&=\sum_{n=1}^{N}\phi_n(\ve{\psi}) \left( \frac{\partial s_{n}(\ve{\theta})}{\partial \ve{\theta}}\right)^{\T}  \frac{\partial s_{n}(\ve{\theta})}{\partial \ve{\theta}}\notag\\
&=\sum_{n=1}^{N}\phi_n(\ve{\psi}) \ve{F}_n(\ve{\theta}),\label{eq:FIMtt}
\end{align}
where the first equality stems from the property of the FIM with independent samples \eqref{eq:FIM:independence}. Accordingly, with
\begin{align}
\frac{\partial \ln p_{{z}_n}(z_{n}|\ve{\psi}) }{\partial {\alpha}} = -  \frac{z_n \exp{ \big(-\frac{z_n(\alpha - s_{n}(\ve{\theta}))^2}{2} \big)} }{\sqrt{2 \pi}\qfunc{z_n(\alpha - s_{n}(\ve{\theta}))} },
\end{align}
we write \eqref{eq:Jat} and \eqref{eq:Jaa} as
\begin{align}
\ve{J}_{\ve{\theta}\alpha}(\ve{\psi})&=
-\sum_{n=1}^{N}\phi_n(\ve{\psi}) \left( \frac{\partial s_{n}(\ve{\theta})}{\partial \ve{\theta}}\right)^{\T}\notag\\
&=-\sum_{n=1}^{N}\phi_n(\ve{\psi}) \ve{f}_n(\ve{\theta}),\label{eq:FIMta}\\
{J}_{\alpha\alpha}(\ve{\psi})&=
\sum_{n=1}^{N}\phi_n(\ve{\psi}).
\label{eq:FIMaa}
\end{align}

In the case where the threshold ${\alpha}$ is known to the receiver, the asymptotic MSE of the MLE 
\begin{align}\label{definition:mle:quantized:genius}
\ve{\hat{\theta}}_{\ve{z}}^{\star}(\ve{z})&\triangleq\arg \max_{\ve{\theta}\in \ve{\Theta}} p_{\ve{z}}(\ve{z}|\ve{\theta},{\alpha})\notag\\
&=\arg \max_{\ve{\theta} \in \ve{\Theta}} \sum_{n=1}^{N} \ln p_{{z}_n}({z}_n|\ve{\theta},{\alpha})
\end{align}
is equivalent to
\begin{align}
\label{eq:CRLB:1bit:genius}
\ve{\text{MSE}}_{\ve{z}}^{\star}(\ve{\psi})\overset{a}{=} \ve{J}_{\ve{\theta}\ve{\theta}}^{-1}(\ve{\psi}).
\end{align}
We will use the results from this subsection to determine the loss induced by hard-limiting with an unknown threshold.
\subsubsection{Performance Measures}
For the comparison between the performance of the ideal \eqref{def:mle:ideal} and the quantized receivers \eqref{definition:mle:quantized} and \eqref{definition:mle:quantized:genius}, we define the average ratios between the MSEs
\begin{align}\label{def:quantization:loss:det}
\chi(\ve{\psi})&\triangleq\frac{1}{K}\sum_{k=1}^{K}\frac{ \big[\text{MSE}_{\ve{y}}(\ve{\theta})\big]_{kk} }{ \big[\text{MSE}_{\ve{z} }(\ve{\psi})\big]_{kk} },\\
\label{def:quantization:loss:det:genius}
\chi^{\star}(\ve{\psi})&\triangleq\frac{1}{K}\sum_{k=1}^{K}\frac{ \big[\text{MSE}_{\ve{y}}(\ve{\theta})\big]_{kk} }{ \big[\text{MSE}^{\star}_{\ve{z} }(\ve{\psi})\big]_{kk} }.
\end{align}
The measures in \eqref{def:quantization:loss:det} and \eqref{def:quantization:loss:det:genius} are chosen since they characterize the performance loss (averaged over the $K$ parameters) which is introduced by hard-limiting and they assure scale invariance. One could choose other ratio-based measures, such as the ratio of average MSEs, which are scale-invariant. However, the ratio of average MSEs may be biased if one of the MSEs is much larger than the others. 

The performance loss introduced in the quantized case by having to estimate the unknown threshold in \eqref{definition:mle:quantized} can be written
\begin{align}\label{def:offset:loss:det}
\Upsilon(\ve{\psi})&\triangleq\frac{1}{K}\sum_{k=1}^{K}\frac{ \big[\text{MSE}^{\star}_{\ve{z}}(\ve{\psi})\big]_{kk} }{ \big[\text{MSE}_{\ve{z} }(\ve{\psi})\big]_{kk} }.
\end{align}
We will use the performance measures from this subsection to quantify the quantization loss in different scenarios.
\subsection{Hybrid Modeling Approach}
As a second approach, we consider the case where the parameter $\ve{\theta}\sim p(\ve{\theta})$ is modeled as a random vector and the threshold ${\alpha}$ as an unknown deterministic nuisance parameter. In this hybrid framework, the errors of the estimators $\ve{\hat{\theta}}_{\ve{y}}(\ve{y})$ and $\ve{\hat{\theta}}_{\ve{z}}(\ve{z})$ are defined
\begin{align}\label{definition:mse:unquantized:receiver:hybrid}
\ve{\text{MSE}}_{\ve{y}}&\triangleq\exdi{\ve{y},\ve{\theta}}{\big(\ve{\hat{\theta}}_{\ve{y}}(\ve{y})-\ve{\theta}\big) \big(\ve{\hat{\theta}}_{\ve{y}}(\ve{y})-\ve{\theta}\big)^{\T}},\\
\label{definition:mse:quantized:receiver:hybrid}
\ve{\text{MSE}}_{\ve{z}}(\alpha)&\triangleq\exdi{\ve{z},\ve{\theta}|\alpha}{\big(\ve{\hat{\theta}}_{\ve{z}}(\ve{z})-\ve{\theta}\big) \big(\ve{\hat{\theta}}_{\ve{z}}(\ve{z})-\ve{\theta}\big)^{\T}}.
\end{align}
\subsubsection{Estimation Procedure}
The asymptotically optimum estimator with the ideal receiver \eqref{system:model:ideal:generic} concerning the performance characterization \eqref{definition:mse:unquantized:receiver:hybrid} is the \textit{maximum a-posteriori probability} (MAP) estimator \cite{Trees07}
\begin{align}\label{eq:MAPy}
\ve{\hat{\theta}}_{\ve{y}}(\ve{y})&\triangleq\arg \max_{\ve{\theta}\in \ve{\Theta}} p_{\ve{y},\ve{\theta}}(\ve{y},\ve{\theta})\notag\\
&=\arg \max_{ \ve{\theta}\in \ve{\Theta} } \big( \ln p_{\ve{y}}(\ve{y}|\ve{\theta})+\ln p_{\ve{\theta}}(\ve{\theta})\big),
\end{align}
where the last equality stems from the Bayes' law. For the $1$-bit receiver \eqref{system:model:sign}, the asymptotically optimum estimator \cite{Bar15_2} concerning \eqref{definition:mse:quantized:receiver:hybrid} is the joint MAP-MLE (JMAP-MLE) \cite{Yeredor00}
\begin{align}\label{eq:MAP:quant}
\begin{bmatrix}\ve{\hat{\theta}}_{\ve{z}}(\ve{z}) &\hat{\alpha}(\ve{z})\end{bmatrix}^{\T}&\triangleq\arg \max_{\ve{\psi}\in \ve{\Psi}} p_{\ve{z}}(\ve{z},\ve{\theta}| {\alpha})\notag\\
&=\arg \max_{\ve{\psi}\in \ve{\Psi}} \big( \ln p_{\ve{z}}(\ve{z}|\ve{\theta}, \alpha)+ \ln p_{\ve{\theta}}(\ve{\theta}) \big),
\end{align}
where the last equality stems from the Bayes' law and the assumption that the prior probability density function of the random parameters $\ve{\theta}$ is independent of the threshold $\alpha$.

For the ideal receiver, the asymptotic performance of the MAP estimator is obtained by using the expected value of the CRLB in (\ref{eq:CRLB}), known as the expected CRLB (ECRLB) \cite{Tabrikian99} \cite[p. 6]{Trees07}, such that the MSE of the optimal infinite-resolution receiver \eqref{definition:mse:unquantized:receiver:hybrid} asymptotically converges to
\begin{align}
\label{eq:ECRLB}
\text{MSE}_{\ve{y}} \overset{a}{=} \exdi{\ve{\theta}}{\ve{F}^{-1}(\boldsymbol{\theta})}.
\end{align}
Note that traditionally, the MSE of Bayesian parameter estimators is lower bounded by the Bayesian CRLB (BCRLB) \cite[p. 5]{Trees07}, given by
\begin{align}
\label{eq:BCRLB}
\text{MSE}_{\ve{y}} \succeq \big(\exdi{\ve{\theta}}{\ve{F}(\boldsymbol{\theta})} +\mathbf{J}_P\big)^{-1},
\end{align}
where the notation $\mathbf{A}\succeq\mathbf{B}$ states that $\mathbf{A}-\mathbf{B}$ is a positive-semidefinite matrix, and $\mathbf{J}_P$ is the prior FIM, given by
\begin{align}
\mathbf{J}_P\triangleq\exdi{\ve{\theta}}{ \bigg(\frac{\partial \ln p_{\ve{\theta}}(\ve{\theta}) }{\partial \ve{\theta}} \bigg)^{\T} \frac{\partial \ln p_{\ve{\theta}}(\ve{\theta}) }{\partial \ve{\theta}} }.
\end{align}
However, this bound is only attainable in special cases, while the ECRLB is in general asymptotically attainable \cite[p. 6]{Trees07}.

For the performance analysis of the $1$-bit receive model \eqref{system:model:sign}, one can suggest the utilization of the hybrid CRLB (HCRLB), given by \cite{HCRLB,Noam,Reuven}
\begin{align}
\label{eq:HCRLB:1bit}
&\ve{\text{MSE}}_{\ve{z}}(\alpha)\succeq\notag\\
&\Bigg( \exdi{\ve{\theta}}{\ve{J}_{\ve{\theta}\ve{\theta}}(\ve{\psi})}-  \frac{\exdi{\ve{\theta}}{\ve{J}_{\ve{\theta}\alpha}(\ve{\psi})}    \exdi{\ve{\theta}}{\ve{J}_{\alpha\ve{\theta}}(\ve{\psi})}}{\exdi{\ve{\theta}}{{J}_{\alpha\alpha} (\ve{\psi})}}  +\mathbf{J}_P  \Bigg)^{-1}.
\end{align}
This bound is traditionally used to lower bound the MSE of unbiased parameter estimators in the hybrid setup. However, in contrast to the CRLB, this lower bound is only attainable in special cases \cite{Bar15_2}. Thus, to characterize the asymptotic performance of the JMAP-MLE, the following theorem is given. This theorem utilizes the definition of the MLE for the hybrid scenario, given by
\begin{align}
\begin{bmatrix}\hat{\ve{\theta}}_{\ve{z}}(\ve{z}) & \hat{\alpha}(\ve{z})\end{bmatrix}^{\T}=\arg \max_{\ve{\psi} \in \ve{\Psi} } \ln p_{\ve{z}}(\ve{z}|\ve{\psi}).
\end{align}
\begin{theorem}[Expected HCRLB (EHCRLB)]\normalfont
	Let us assume the following regularity conditions:
	\begin{enumerate}
		\item The solution of the JMAP-MLE converges to the solution of the MLE in probability \cite{Ibragimov81} as $N$ tends to infinity.
		\item The sequence of MLEs as a function of the number of measurements is asymptotically uniformly integrable \cite{Jeganathan82}. 
	\end{enumerate}
	Then, 
	\begin{equation}
	\label{eq:EHCRLB}
	\text{MSE}_{\ve{z}}(\alpha)\overset{a}{=}\exdi{\ve{\theta}}{\bigg( \ve{J}_{\ve{\theta}\ve{\theta}}(\ve{\psi}) -  \frac{\ve{J}_{\ve{\theta}\alpha}(\ve{\psi})    \ve{J}_{\alpha\ve{\theta}}(\ve{\psi})}{{J}_{\alpha\alpha} (\ve{\psi})}   \bigg)^{-1}}.
	\end{equation}
\end{theorem}
\begin{IEEEproof}
see Appendix \ref{app:EHCRLB}.
\end{IEEEproof}
Note that the r.h.s. of \eqref{eq:EHCRLB} represents the hybrid version of the ECRLB, denoted by EHCRLB. While it does not constitute a lower bound, the EHCRLB is asymptotically attainable by the JMAP-MLE. To the best of the authors' knowledge, no previous work in the literature has presented this performance analysis tool in the hybrid context. Again, due to Jensen's inequality \cite[p. 83-84]{Inequalities88}
\begin{align}\label{eq:HCRB:1bit}
&\exdi{\ve{\theta}}{\bigg( \ve{J}_{\ve{\theta}\ve{\theta}}(\ve{\psi}) -  \frac{\ve{J}_{\ve{\theta}\alpha}(\ve{\psi})    \ve{J}_{\alpha\ve{\theta}}(\ve{\psi})}{{J}_{\alpha\alpha} (\ve{\psi})}   \bigg)^{-1}}\succeq\notag\\
&\succeq\Bigg\{\exdi{\ve{\theta}}{\bigg( \ve{J}_{\ve{\theta}\ve{\theta}}(\ve{\psi}) -  \frac{\ve{J}_{\ve{\theta}\alpha}(\ve{\psi})    \ve{J}_{\alpha\ve{\theta}}(\ve{\psi})}{{J}_{\alpha\alpha} (\ve{\psi})}   \bigg)}\Bigg\}^{-1}\notag\\
&\succeq\Bigg\{ \exdi{\ve{\theta}}{\ve{J}_{\ve{\theta}\ve{\theta}}(\ve{\psi})} -  \frac{\exdi{\ve{\theta}}{\ve{J}_{\ve{\theta}\alpha}(\ve{\psi})}    \exdi{\ve{\theta}}{\ve{J}_{\alpha\ve{\theta}}(\ve{\psi})}}{\exdi{\ve{\theta}}{{J}_{\alpha\alpha} (\ve{\psi})}}   \Bigg\}^{-1},
\end{align}
where the last inequality is obtained using the covariance inequality \cite[p. 113]{LehmannBook98}, given by
\begin{equation}
\exdi{\ve{\theta}}{\ve{u}\ve{u}^{\T}}\succeq\exdi{\ve{\theta}}{\ve{u}\ve{w}^{\T}}\exdiinv{\ve{\theta}}{\ve{w}\ve{w}^{\T}}\exdi{\ve{\theta}}{\ve{w}\ve{u}^{\T}},
\end{equation}
for some random vectors $\ve{u}$ and $\ve{w}$, by setting $\ve{u}=\frac{\ve{J}_{\ve{\theta}\alpha}(\ve{\psi})    }{\sqrt{{J}_{\alpha\alpha} (\ve{\psi})}}$ and $w=\sqrt{{J}_{\alpha\alpha} (\ve{\psi})}$. The r.h.s. of \eqref{eq:HCRB:1bit} can be identified as the asymptotic version (when the prior information about $\ve{\theta} $ is negligible) of the HCRLB in \eqref{eq:HCRLB:1bit}. That is, while the expression in \eqref{eq:EHCRLB} can in general be asymptotically achieved by the JMAP-MLE, the r.h.s. of \eqref{eq:HCRB:1bit} serves only as a lower bound and is only achieved under special conditions \cite{Bar15_2}. Thus, the EHCRLB and the HCRLB present relations similar to the aforementioned relations between the ECRLB and BCRLB. 

In case that the quantization offset $\alpha$ is known to the receiver, we proceed by using the MAP estimator
\begin{align}\label{eq:MAP:quant:genius}
\hat{\ve{\theta}}^{\star}(\ve{z}) =\arg \max_{\ve{\theta} \in \ve{\Theta} } \ln p_{\ve{z}}(\ve{z}|\ve{\theta},\alpha)
\end{align}
with asymptotic MSE
\begin{equation}
\label{eq:EHCRLB:genius}
\text{MSE}_{\ve{z}}^{\star}(\alpha)\overset{a}{=}\exdi{\ve{\theta}}{\ve{J}_{\ve{\theta}\ve{\theta}}^{-1}(\ve{\psi}) }
\end{equation}
and the error bound
\begin{align}
\label{eq:HCRLB:1bit:genius}
\ve{\text{MSE}}_{\ve{z}}^{\star}(\alpha)&\succeq\Big( \exdi{\ve{\theta}}{\ve{J}_{\ve{\theta}\ve{\theta}}(\ve{\psi})}+\mathbf{J}_P \Big)^{-1}.
\end{align}
Like for the deterministic case we use the results from this subsection to determine the hard-limiting loss.
\subsubsection{Performance Measures}
Note that for the hybrid modeling approach the quantization loss measures
\begin{align}\label{def:quantization:loss:hyp}
\chi(\alpha)&\triangleq\frac{1}{K}\sum_{k=1}^{K}\frac{ \big[\text{MSE}_{\ve{y}}\big]_{kk} }{ \big[\text{MSE}_{\ve{z} }(\alpha)\big]_{kk} },\\
\label{def:quantization:loss:hyp:genius}
\chi^{\star}(\alpha)&\triangleq\frac{1}{K}\sum_{k=1}^{K}\frac{ \big[\text{MSE}_{\ve{y}}\big]_{kk} }{ \big[\text{MSE}^{\star}_{\ve{z} }(\alpha)\big]_{kk} },
\end{align}
only depend on the quantization offset $\alpha$. According to the deterministic modeling approach, we define the performance penalty introduced by the estimation of the unknown quantization offset as
\begin{align}\label{def:loss:offset:hyp}
\Upsilon(\alpha)&\triangleq\frac{1}{K}\sum_{k=1}^{K}\frac{ \big[\text{MSE}^{\star}_{\ve{z}}(\alpha)\big]_{kk} }{ \big[\text{MSE}_{\ve{z} }(\alpha)\big]_{kk} }.
\end{align}

The generic expressions derived in this section will be used in the following to obtain analytic insights or to conduct numerical evaluations and simulations. 
\section{Performance Analysis for Low SNR}\label{sec:lowSNR}
In this section, we establish the conditions for which the loss due to the missing knowledge of the offset vanishes. To this end, the generic results of the deterministic and the hybrid approach are discussed under the assumption that the channel estimation task takes place in the low SNR regime. Such an assumption is well-motivated in cases where the radio transmitter and receiver are far apart, like for example in a satellite communication link or when weak receive signals have to be processed as in radar applications. To define the low SNR regime consistently, we assume the existence of some (not necessarily unique)  $\ve{\theta}_0\in\boldsymbol{\Theta}$ for which 
\begin{align}\label{eq:lowSNR}
{s}_n( \ve{\theta})\to 0,\ \forall n, \text{ when } \ve{\theta} \to \ve{\theta}_0.
\end{align}
We will use the limiting case \eqref{eq:lowSNR} of low SNR to evaluate the performance of the algorithms discussed in Sec. \ref{sec:Bound} and to identify favorable conditions on the derivative $\ve{f}_n(\ve{\theta})$ of the pilot signal ${s}_n( \ve{\theta})$ under an unknown quantization threshold.
\subsection{Deterministic Approach}
Since in the low SNR regime the pilot signal ${s}_n( \ve{\theta})$ tends to zero, we define
\begin{align}\label{definition:phi:zero}
\phi_0(\alpha)&\triangleq\lim_{ \ve{\theta} \to \ve{\theta}_0 }\phi_n(\ve{\psi})\notag\\
&=\frac{ \expb{ -\alpha ^2 } }{ {2\pi} \big(\qfunc{\alpha}-\qfuncptwo{\alpha}\big)}.
\end{align}
Hence, with the functions $\ve{F}(\ve{\theta}), \ve{F}_n (\ve{\theta}), \ve{f}_n(\ve{\theta})$ defined in \eqref{info:fishery}, \eqref{info:fishery:sample}, \eqref{eq:smallFIM:sample} and
\begin{align}\label{eq:smallFIM}
\ve{f}(\ve{\theta})&\triangleq\sum_{n=1}^{N}\ve{f}_n(\ve{\theta}),
\end{align}
the FIM elements in \eqref{eq:FIMtt}, \eqref{eq:FIMta}, and \eqref{eq:FIMaa} associated with the quantized receiver, become
\begin{align}\label{lowsnr:FIMtt}
\lim_{ \ve{\theta} \to \ve{\theta}_0 } \ve{J}_{\ve{\theta}\ve{\theta}}(\ve{\psi}) &= \sum_{n=1}^{N}\phi_0(\alpha)\ve{F}_n (\ve{\theta}_0)\notag\\
&=\phi_0(\alpha) \ve{F}(\ve{\theta}_0),\\
\label{lowsnr:FIMta}
\lim_{ \ve{\theta} \to \ve{\theta}_0 } \ve{J}_{\ve{\theta}\alpha}(\ve{\psi}) &= -\sum_{n=1}^{N}\phi_0(\alpha)\ve{f}_n(\ve{\theta}_0)\notag\\
&=-\phi_0(\alpha) \ve{f}(\ve{\theta}_0),
\end{align}
and
\begin{align}\label{lowsnr:FIMaa}
\lim_{ \ve{\theta} \to \ve{\theta}_0 }J_{\alpha\alpha}(\ve{\psi})&=N\phi_0(\alpha).
\end{align}
Substitution of \eqref{lowsnr:FIMtt}-\eqref{lowsnr:FIMaa} into \eqref{eq:CRLB:1bit}, yields
\begin{align}\label{lowsnr:CRLB:1bit}
\lim_{ \ve{\theta} \to \ve{\theta}_0 }\ve{\text{MSE}}_{\ve{z}}(\ve{\psi}\nonumber )&\overset{a}{=}\lim\limits_{\ve{\theta} \to \ve{\theta}_0}\bigg( \ve{J}_{\ve{\theta}\ve{\theta}}(\ve{\psi}) -  \frac{\ve{J}_{\ve{\theta}\alpha}(\ve{\psi})\ve{J}_{\alpha\ve{\theta}}(\ve{\psi})}{{J}_{\alpha\alpha} (\ve{\psi})}\bigg)^{-1}\\&=\frac{1}{\phi_0(\alpha)}\Bigg(\ve{F}(\ve{\theta}_0) - \frac{1}{N}\ve{f}(\ve{\theta}_0) \ve{f}^{\T}(\ve{\theta}_0)  \Bigg)^{-1}.
\end{align}

From \eqref{eq:FIM:independence} and \eqref{eq:smallFIM} it can be observed that the entries of $\ve{F}(\ve{\theta})$ and $\ve{f}(\ve{\theta})$ grow linearly with the number of samples $N$. For all cases where, due to the channel model or the pilot signal design, the matrix entries of
\begin{align}
\ve{f}(\ve{\theta}) \ve{f}^{\T}(\ve{\theta}) = \sum_{n=1}^{N} \ve{f}_n(\ve{\theta}) \ve{f}_n^{\T}(\ve{\theta}) + \sum_{\substack{n=1, m=1\\ n \neq m}}^{N,N} \ve{f}_n(\ve{\theta}) \ve{f}_m^{\T}(\ve{\theta}) 
\end{align}
exhibit a growth rate of linear order, i.e.,
\begin{align}\label{linear:scale:assumption:pilot}
\ve{f}(\ve{\theta}) \ve{f}^{\T}(\ve{\theta}) \sim \mathcal{O}(N),
\end{align}
the asymptotic $1$-bit MSE in the low SNR regime $\eqref{lowsnr:CRLB:1bit}$ becomes
\begin{align}\label{lowsnr:CRLB:1bit:simplified}
\lim_{ \ve{\theta} \to \ve{\theta}_0 }\ve{\text{MSE}}_{\ve{z}}(\ve{\psi})&\overset{a}{=} \frac{1}{\phi_0(\alpha)}\ve{F}^{-1}(\ve{\theta}_0).
\end{align}
Note that in \eqref{lowsnr:CRLB:1bit:simplified} we use the fact that in the expression \eqref{lowsnr:CRLB:1bit} the FIM $\ve{F}(\ve{\theta})$ grows linearly in $N$ while with the condition \eqref{linear:scale:assumption:pilot} the term $\frac{1}{N}\ve{f}(\ve{\theta}) \ve{f}^{\T}(\ve{\theta})$ stays constant, such that in the asymptotic regime $\ve{F}(\ve{\theta})$ dominates the r.h.s. of \eqref{lowsnr:CRLB:1bit}. 

With the low SNR performance of the ideal receive system
\begin{align}\label{lowsnr:CRLB:inftybit}
\lim_{ \ve{\theta} \to \ve{\theta}_0 } \text{MSE}_{\ve{y}}(\ve{\theta})\overset{a}{=}\ve{F}^{-1}(\ve{\theta}_0),
\end{align}
the quantization loss \eqref{def:quantization:loss:det} then tends towards
\begin{align}\label{def:quantization:loss:det:low:snr}
\lim_{ \ve{\theta} \to \ve{\theta}_0 }\chi(\ve{\psi})&=\frac{1}{K}\sum_{k=1}^{K} \frac{ \big[\lim_{ \ve{\theta} \to \ve{\theta}_0 } \text{MSE}_{\ve{y}}(\ve{\theta})\big]_{kk} }{ \big[ \lim_{ \ve{\theta} \to \ve{\theta}_0 } \text{MSE}_{\ve{z} }(\ve{\psi})\big]_{kk} }\notag\\
&\overset{a}{=}\phi_0(\alpha).
\end{align}
Note, that for the symmetric case, i.e., $\alpha=0$, with $\qfunc{0}=\frac{1}{2}$ we obtain the classical coarse quantization result \cite{Vleck66}
\begin{align}
\phi_0(0)=\frac{ 1}{ {2\pi} \big(\qfunc{0}-\qfuncptwo{0}\big)}=\frac{2}{\pi}.
\end{align}
For the quantized receiver with known offset \eqref{definition:mle:quantized:genius}, we have
\begin{align}\label{lowsnr:CRLB:1bit:genius}
\lim_{ \ve{\theta} \to \ve{\theta}_0 }\ve{\text{MSE}}_{\ve{z}}^{\star}(\ve{\psi} )&\overset{a}{=}\frac{1}{\phi_0(\alpha)}\ve{F}^{-1}(\ve{\theta}_0)
\end{align}
and
\begin{align}\label{def:quantization:loss:det:genius:low:snr}
\lim_{ \ve{\theta} \to \ve{\theta}_0 }\chi^{\star}(\ve{\psi})&=\frac{1}{K}\sum_{k=1}^{K} \frac{ \big[\lim_{ \ve{\theta} \to \ve{\theta}_0 } \text{MSE}_{\ve{y}}(\ve{\theta})\big]_{kk} }{ \big[ \lim_{ \ve{\theta} \to \ve{\theta}_0 } \text{MSE}_{\ve{z}}^{\star}(\ve{\psi})\big]_{kk} }\notag\\
&\overset{a}{=}\phi_0(\alpha),
\end{align}
such that if \eqref{linear:scale:assumption:pilot} is satisfied, the loss \eqref{def:offset:loss:det} introduced by the estimation of the unknown offset in \eqref{definition:mle:quantized} vanishes
\begin{align}\label{offset:estimation:loss:lowSNR}
\lim_{ \ve{\theta} \to \ve{\theta}_0 }\Upsilon(\ve{\psi})&\overset{a}{=}1.
\end{align}
\subsection{Hybrid Approach}
\label{ssec:hybrid:lowSNR}
In this subsection, we provide an analysis similar to the deterministic one for the hybrid modeling approach. To adapt the low SNR regime definition \eqref{eq:lowSNR} to this scenario, we interpret the required limit procedure in the following manner. It is assumed that the prior $p_{\ve{\theta}}(\ve{\theta})$ can be controlled by a set of parameters $\ve{\gamma}\in\ve{\Gamma}$, such that $p_{\ve{\theta}}(\ve{\theta};\ve{\gamma})$ stands for the parameterized prior. Furthermore,
\begin{align}\label{eq:hybrid:lowSNR}
\exists\ve{\gamma}_0:\ \lim\limits_{\ve{\gamma} \to \ve{\gamma}_0}p_{\ve{\theta}}({s}_n( \ve{\theta})= 0;\ve{\gamma})=1,\ \forall n,\ve{\theta}\in\ve{\Theta}(\ve{\gamma}_0),
\end{align}
where
\begin{align}
\ve{\Theta}(\ve{\gamma}_0)=\big\{ \ve{\theta} \, | \lim_{ \ve{\gamma} \to \ve{\gamma}_0 } p_{\ve{\theta}}(\ve{\theta};\ve{\gamma}) \neq 0 \big\}
\end{align}
is the significant support of the random parameter $\ve{\theta}$ at $\ve{\gamma}_0$.
By taking the limit of both sides of \eqref{eq:EHCRLB} as $\ve{\gamma}$ tends to $\ve{\gamma}_0$ yields
\begin{align}\label{eq:hybrid:lowSNR:limit}
&\lim_{ \ve{\gamma} \to \ve{\gamma}_0} \text{MSE}_{\ve{z}}(\alpha)\notag\\
&\overset{a}{=}\lim_{ \ve{\gamma} \to \ve{\gamma}_0}\exdi{\ve{\theta};\ve{\gamma}}{\bigg( \ve{J}_{\ve{\theta}\ve{\theta}}(\ve{\psi}) -  \frac{\ve{J}_{\ve{\theta}\alpha}(\ve{\psi})\ve{J}_{\alpha\ve{\theta}}(\ve{\psi})}{{J}_{\alpha\alpha} (\ve{\psi})}\bigg)^{-1}}.
\end{align}
Hence, under the assumption that the CRLB, $\bigg( \ve{J}_{\ve{\theta}\ve{\theta}}(\ve{\psi}) -  \frac{\ve{J}_{\ve{\theta}\alpha}(\ve{\psi})\ve{J}_{\alpha\ve{\theta}}(\ve{\psi})}{{J}_{\alpha\alpha} (\ve{\psi})}\bigg)^{-1}$ for estimating $\ve{\theta}$ with unknown offset $\alpha$ is uniformly bounded in the vicinity of $\boldsymbol{\gamma}_0$ for any fixed $N$ and with $p_{\ve{z}}(\ve{z},\ve{\theta}| {\alpha};\boldsymbol{\gamma})\leq1$, the uniform convergence theorem \cite{Vaart} and \eqref{eq:hybrid:lowSNR} imply that 
\begin{align}\label{eq:hybrid:lowSNR:limit2}
&\lim_{ \ve{\gamma} \to \ve{\gamma}_0}\exdi{\ve{\theta};\ve{\gamma}}{\bigg( \ve{J}_{\ve{\theta}\ve{\theta}}(\ve{\psi}) -  \frac{\ve{J}_{\ve{\theta}\alpha}(\ve{\psi})\ve{J}_{\alpha\ve{\theta}}(\ve{\psi})}{{J}_{\alpha\alpha} (\ve{\psi})}\bigg)^{-1}}\notag\\
&=\lim_{ \ve{\gamma} \to \ve{\gamma}_0}\exdi{\ve{\theta};\ve{\gamma}}{\lim_{\underset{\forall n}{ s_n(\boldsymbol{\theta}) \to 0}}\bigg( \ve{J}_{\ve{\theta}\ve{\theta}}(\ve{\psi}) -  \frac{\ve{J}_{\ve{\theta}\alpha}(\ve{\psi})\ve{J}_{\alpha\ve{\theta}}(\ve{\psi})}{{J}_{\alpha\alpha} (\ve{\psi})}\bigg)^{-1}}
\notag\\&=\lim_{ \ve{\gamma} \to \ve{\gamma}_0}\exdi{\ve{\theta};\ve{\gamma}}{\frac{1}{\phi_0(\alpha)}\Bigg(\ve{F}(\ve{\theta}) - \frac{1}{N}\ve{f}(\ve{\theta}) \ve{f}^{\T}(\ve{\theta})  \Bigg)^{-1}}
\notag\\
&\overset{a}{=}\frac{1}{\phi_0(\alpha)}\exdi{\ve{\theta};\ve{\gamma}_0}{\ve{F}^{-1}(\ve{\theta})},
\end{align}
where the second equality follows the last equality in \eqref{lowsnr:CRLB:1bit} and the asymptotic equality stems from the assumption \eqref{linear:scale:assumption:pilot}. Finally,
\begin{align}\label{eq:SNR:hybrid:preMSE}
\lim_{ \ve{\gamma} \to \ve{\gamma}_0} \text{MSE}_{\ve{z}}(\alpha)&\overset{a}{=}\lim_{ \ve{\gamma} \to \ve{\gamma}_0} \text{MSE}_{\ve{z}}^{\star}(\alpha)\notag\\
&=\frac{1}{\phi_0(\alpha)}\exdi{\ve{\theta};\ve{\gamma}_0}{\ve{F}^{-1}(\ve{\theta})}
\end{align}
for all cases where \eqref{linear:scale:assumption:pilot} holds, while \eqref{eq:ECRLB} leads to
\begin{align}\label{eq:SNR:hybrid:MSE}
\lim_{ \ve{\gamma} \to \ve{\gamma}_0} \text{MSE}_{\ve{y}}\overset{a}{=}
&\exdi{\ve{\theta};\ve{\gamma}_0}{\ve{F}^{-1}(\ve{\theta})}.
\end{align}
Therefore, under the restriction in \eqref{linear:scale:assumption:pilot}, for the hybrid quantization losses \eqref{def:quantization:loss:hyp} and \eqref{def:quantization:loss:hyp:genius},
\begin{align}\label{def:quantization:loss:hyp:low:snr}
\lim_{ \ve{\gamma} \to \ve{\gamma}_0} \chi(\alpha)&\overset{a}{=}\lim_{ \ve{\gamma} \to \ve{\gamma}_0} \chi^{\star}(\alpha)=\phi_0(\alpha),
\end{align}
such that the accuracy loss due to offset estimation in \eqref{def:loss:offset:hyp} vanishes when operating in the low SNR regime, i.e.,
\begin{align}\label{offset:estimation:loss:hyp:lowSNR}
\lim_{ \ve{\gamma} \to \ve{\gamma}_0}
 \Upsilon(\alpha)&\overset{a}{=}1.
\end{align} 
\section{Application - Wireless Channel Estimation}\label{sec:sim}
Using the generic expressions from the previous sections, we analyze the performance gap between the ideal receiver \eqref{system:model:ideal:generic} with high ADC resolution and the low-complexity receiver \eqref{system:model:sign} with $1$-bit ADC for a wireless channel with inter-symbol interference (ISI). Such a channel estimation problem occurs in the application of mobile communication, where channel characteristics like multi-path propagation or nonlinear frequency response of the time-varying wireless propagation medium have to be measured in a recurrent manner.
\subsection{Multi-tap SISO Channel Estimation}\label{sec:siso:multi:tap}
The signal model of the ISI channel is 
\begin{align}\label{def:ISI:ideal:receiver}
y_n = \sum_{k=1}^{K} h_k x_{n-k+1} + \eta_n,
\end{align}
where $h_k\in\fieldR$ is the receive strength of the $k$-th channel tap and ${x}_n\in \{-1,1\}$ a binary pilot signal (BPSK) of known structure, even length $N$ and with symmetric symbol assignment, i.e., $\sum_{n=1}^{N}{x}_n=0$. Further, we define the vector $\ve{x}_n\in \{-1, 1\}^{K}$ with column entries
\begin{align}
\lbrack \ve{x}_n \rbrack_{i} = x_{n-i+1},\quad i=1,\ldots,K
\end{align}
and the matrix $\ve{X}_n\in \{-1, 1\}^{K\times K}$ 
\begin{align}
\ve{X}_n = \ve{x}_n \ve{x}_n^{\T}.
\end{align}
The ISI-channel estimation task is to determine the channel coefficients, summarized in the parameter vector
\begin{align}
\ve{\theta} = \begin{bmatrix} h_1 &h_2 &\ldots &h_{K} \end{bmatrix}^{\T},
\end{align}
from the receive signals 
\begin{align}\label{def:ISI:ideal:receiver:vector}
y_n&={s}_n(\ve{\theta}) + \eta_n=\ve{x}_n^{\T}\ve{\theta} + \eta_n.
\end{align}
A wireless receiver with a $1$-bit A/D conversion observes the quantized signal samples
\begin{align}\label{def:ISI:quantized:receiver}
z_n = \sign{ \ve{x}_n^{\T}\ve{\theta} + \eta_n - \alpha}.
\end{align}
Note that for the considered ISI scenario \eqref{def:ISI:ideal:receiver:vector} one obtains $\ve{f}_n(\ve{\theta})=\ve{x}_n$. Therefore, with a binary pilot signal following a symmetric symbol assignment it can be verified
\begin{align}
\sum_{\substack{n=1, m=1\\ n \neq m}}^{N,N} \ve{x}_n \ve{x}_m^{\T}\sim\mathcal{O}(1),
\end{align}
such that \eqref{linear:scale:assumption:pilot} is fulfilled and the analytic low SNR results \eqref{def:quantization:loss:det:low:snr}, \eqref{offset:estimation:loss:lowSNR}, \eqref{def:quantization:loss:hyp:low:snr}, and \eqref{offset:estimation:loss:hyp:lowSNR} hold for the ISI channel model \eqref{def:ISI:quantized:receiver}.
\subsubsection{Performance Analysis - Deterministic Approach}
Under the deterministic framework the FIM \eqref{info:fishery} for the ideal wireless receive system \eqref{def:ISI:ideal:receiver} is given by
\begin{align}\label{def:ISI:fim:ideal:receiver}
\ve{F} (\ve{\theta})&=  \sum_{n=1}^{N} \ve{X}_n,
\end{align}
such that
\begin{align}
\ve{\text{MSE}}_{\ve{y}}(\ve{\theta} )&\overset{a}{=}\left(\sum_{n=1}^{N}  \ve{X}_n\right)^{-1}.
\end{align}
For the $1$-bit quantized receiver \eqref{def:ISI:quantized:receiver}, with \eqref{definition:phin} we obtain
\begin{multline}
\phi_{n}(\ve{\psi})=\\\frac{ \expb{ -\big(\alpha-\sum_{k}^{} h_k x_{n-k+1}\big)^2 } }{ {2\pi} \Big( \qfunc{\alpha - \sum_{k}^{} h_k x_{n-k+1} } - \qfuncptwo{\alpha- \sum_{k}^{} h_k x_{n-k+1} } \Big) },
\end{multline} 
such that the quantized FIMs \eqref{eq:FIMtt}, \eqref{eq:FIMta}, and \eqref{eq:FIMaa} are
\begin{align}\label{FIM:theta:theta}
\ve{J}_{\ve{\theta}\ve{\theta}}(\ve{\psi}) &=\sum_{n=1}^{N} \phi_{n}(\ve{\psi}) \ve{X}_n,\\
\label{FIM:theta:alpha}
\ve{J}_{\ve{\theta}\alpha}(\ve{\psi})&=-\sum_{n=1}^{N} \phi_{n}(\ve{\psi}) \ve{x}_n,\\
\label{FIM:alpha:alpha}
J_{\alpha\alpha}(\ve{\psi})&=\sum_{n=1}^{N} \phi_{n}(\ve{\psi}).
\end{align}
Under the low SNR assumption, with \eqref{definition:phi:zero} we derive
\begin{align}
\lim_{\ve{\theta} \to \ve{0}} \ve{J}_{\ve{\theta}\ve{\theta}}(\ve{\psi}) &= \phi_0(\alpha) \sum_{n=1}^{N}  \ve{X}_n,\\
\lim_{\ve{\theta} \to \ve{0}}  \ve{J}_{\ve{\theta}\alpha}(\ve{\psi}) &= - \phi_0(\alpha) \sum_{n=1}^{N}  \ve{x}_n,
\end{align}
and
\begin{align}
\lim_{\ve{\theta} \to \ve{0}} J_{\alpha\alpha}(\ve{\psi})&=N \phi_0(\alpha).
\end{align}
With \eqref{lowsnr:CRLB:1bit} and \eqref{lowsnr:CRLB:1bit:genius}, we obtain the asymptotic MSEs
\begin{align}\label{eq:SNR:det:MSE}
\lim_{\ve{\theta} \to \ve{0}}  \ve{\text{MSE}}_{\ve{z}}(\ve{\psi} )&\overset{a}{=}\frac{1}{\phi_0(\alpha)}\left(\sum_{n=1}^{N}  \ve{X}_n\right)^{-1},
\end{align}
and
\begin{align}
\lim_{\ve{\theta} \to \ve{0}}  \ve{\text{MSE}}_{\ve{z}}^{\star}(\ve{\psi} )&\overset{a}{=}\frac{1}{\phi_0(\alpha)}\left(\sum_{n=1}^{N}  \ve{X}_n\right)^{-1},
\end{align}
in the low SNR regime. Therefore, like predicted in \eqref{def:quantization:loss:det:low:snr}, the loss \eqref{def:offset:loss:det} introduced by the unknown offset vanishes, i.e.,
\begin{align}
\lim_{\ve{\theta} \to \ve{0}}\chi(\ve{\psi})&=\lim_{\ve{\theta} \to \ve{0}}\chi^{\star}(\ve{\psi})=\phi_0(\alpha),
\end{align}
in accordance with \eqref{offset:estimation:loss:lowSNR}.
\subsubsection{Results - Deterministic Approach}
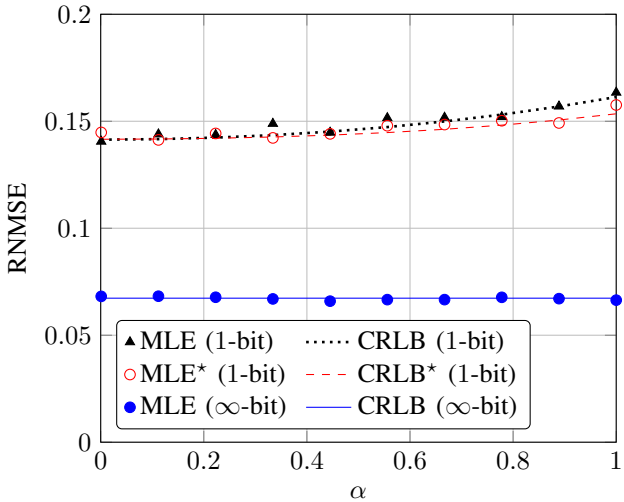
\begin{figure}[!htbp]
\begin{tikzpicture}[scale=1.0]

  	\begin{axis}[ylabel=$\text{RNMSE}$,
  			xlabel=$\alpha$,
			grid,
			tick label style={/pgf/number format/fixed},
			ymin=0.0,
			ymax=0.2,
			xmin=0,
			xmax=1,
			legend columns=2,
			legend style={/tikz/column 2/.style={column sep=5pt}},
			legend pos= south west]
			
			\addplot[black, only marks, mark=triangle*] table[x index=0, y index=1]{Algorithm_ML_ISI3_1bit_m3dB.txt};
			\addlegendentry{MLE ($1$-bit)};
			
			\addplot[black, style=dotted, line width=1pt] table[x index=0, y index=2]{Algorithm_ML_ISI3_1bit_m3dB.txt};
			\addlegendentry{CRLB ($1$-bit)};
			
			\addplot[red, only marks, mark=o] table[x index=0, y index=1]{Algorithm_ML_ISI3_1bit_genius_m3dB.txt};
			\addlegendentry{MLE$^{\star}$ ($1$-bit)};
			
			\addplot[red, style=dashed] table[x index=0, y index=2]{Algorithm_ML_ISI3_1bit_genius_m3dB.txt};
			\addlegendentry{CRLB$^{\star}$ ($1$-bit)};
			
			\addplot[blue, only marks, mark=*] table[x index=0, y index=1]{Algorithm_ML_ISI3_ideal_m3dB.txt};
			\addlegendentry{MLE ($\infty$-bit)};
			
			\addplot[blue] table[x index=0, y index=2]{Algorithm_ML_ISI3_ideal_m3dB.txt};
			\addlegendentry{CRLB ($\infty$-bit)};
			
	\end{axis}

\end{tikzpicture}
\caption{MSE - Deterministic ISI Channel ($\text{SNR}_{1}=-3$ dB)}
\label{fig:PerformanceMLE:ISI:mediumSNR}
\end{figure}

For the simulations of the ISI channel estimation task, we assume
\begin{align}
h_k^2={\text{SNR}_k}.
\end{align}
Considering a scenario with $K=3$ channel taps and $N=1024$ symbols, we set the signal strength of the interfering symbols to $\text{SNR}_2=\text{SNR}_1-3\text{ dB}$, $\text{SNR}_3=\text{SNR}_1-6\text{ dB}$ and average the estimation error of $\ve{\hat{\theta}}_{\ve{y}}(\ve{y})$, $\ve{\hat{\theta}}_{\ve{z}}(\ve{z})$, and $\ve{\hat{\theta}}_{\ve{z}}^{\star}(\ve{z})$ over $1000$ noise realizations. The performance is evaluated by the root-normalized MSE (RNMSE)
\begin{align}\label{definition:RNMSE:infty}
\text{RNMSE}_{\ve{y}}(\ve{\theta})=\sqrt{\frac{1}{K}\sum_{k=1}^{K} \frac{\big[ \text{MSE}_{\ve{y}}(\ve{\theta}) \big]_{kk}}{h_k^2}}
\end{align}
for the ideal receiver and 
\begin{align}
\text{RNMSE}_{\ve{z}}(\ve{\psi})&=\sqrt{\frac{1}{K}\sum_{k=1}^{K} \frac{\big[ \text{MSE}_{\ve{z}}(\ve{\psi}) \big]_{kk}}{h_k^2}},\label{definition:RNMSE:1bit}\\
\text{RNMSE}_{\ve{z}}^{\star}(\ve{\psi})&=\sqrt{\frac{1}{K}\sum_{k=1}^{K} \frac{\big[ \text{MSE}_{\ve{z}}^{\star}(\ve{\psi}) \big]_{kk}}{h_k^2}}\label{definition:RNMSE:1bit:genius}
\end{align}
for the two $1$-bit receive systems.
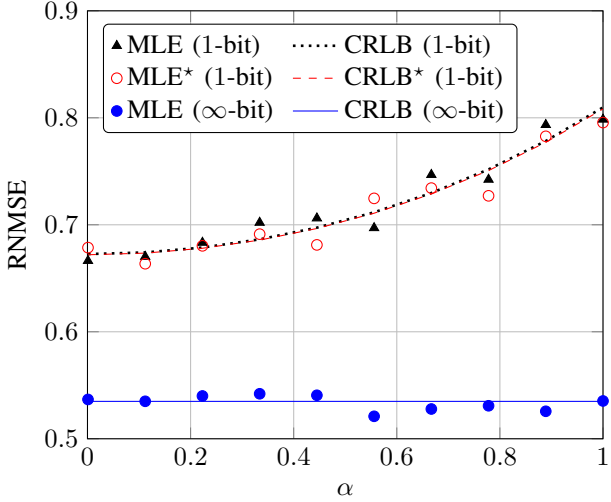
\begin{figure}[!htbp]
\begin{tikzpicture}[scale=1.0]

  	\begin{axis}[ylabel=$\text{RNMSE}$,
  			xlabel=$\alpha$,
			grid,
			ymin=0.5,
			ymax=0.9,
			xmin=0,
			xmax=1,
			legend columns=2,
			legend style={/tikz/column 2/.style={column sep=5pt}},
			legend pos= north west]
						
			\addplot[black, only marks, mark=triangle*] table[x index=0, y index=1]{Algorithm_ML_ISI3_1bit_m21dB.txt};
			\addlegendentry{MLE ($1$-bit)};
			
			\addplot[black, style=dotted, line width=1pt] table[x index=0, y index=2]{Algorithm_ML_ISI3_1bit_m21dB.txt};
			\addlegendentry{CRLB ($1$-bit)};
			
			\addplot[red, only marks, mark=o] table[x index=0, y index=1]{Algorithm_ML_ISI3_1bit_genius_m21dB.txt};
			\addlegendentry{MLE$^{\star}$ ($1$-bit)};
			
			\addplot[red, style=dashed] table[x index=0, y index=2]{Algorithm_ML_ISI3_1bit_genius_m21dB.txt};
			\addlegendentry{CRLB$^{\star}$ ($1$-bit)};
			
			\addplot[blue, only marks, mark=*] table[x index=0, y index=1]{Algorithm_ML_ISI3_ideal_m21dB.txt};
			\addlegendentry{MLE ($\infty$-bit)};
			
			\addplot[blue] table[x index=0, y index=2]{Algorithm_ML_ISI3_ideal_m21dB.txt};
			\addlegendentry{CRLB ($\infty$-bit)};
			
	\end{axis}

\end{tikzpicture}
\caption{MSE - Deterministic ISI Channel ($\text{SNR}_{1}=-21$ dB)}
\label{fig:PerformanceMLE:ISI:lowSNR}
\end{figure}
Figs. \ref{fig:PerformanceMLE:ISI:lowSNR} and \ref{fig:PerformanceMLE:ISI:mediumSNR} illustrates the RNMSEs \eqref{definition:RNMSE:infty}-\eqref{definition:RNMSE:1bit:genius} for the low SNR regime ($\text{SNR}_1=-21$ dB) and the medium SNR regime ($\text{SNR}_1=-3$ dB), respectively.
It can be observed that for both scenarios (Figs. \ref{fig:PerformanceMLE:ISI:lowSNR}  and \ref{fig:PerformanceMLE:ISI:mediumSNR}) the CRLBs accurately characterize the performance of the MLEs.
\begin{figure}[!htbp]
\begin{tikzpicture}[scale=1.0]

  	\begin{axis}[ylabel=$\chi\text{ [dB]}$,
  			xlabel=$\alpha$,
			grid,
			tick label style={/pgf/number format/fixed},
			ymin=-6.0,
			ymax=-1.0,
			xmin=0,
			xmax=1,
			legend columns=2,
			legend style={/tikz/column 2/.style={column sep=5pt}},
			legend pos= south west]
			
    			\addplot[black, style=solid, line width=0.75pt,smooth,every mark/.append style={solid}, mark=otimes*, mark repeat=1] table[x index=0, y index=1]{HardLimLoss_ML_ISI3_m21dB.txt};
			\addlegendentry{$\chi\phantom{^{\star}} (-21\text{ dB})$};
			
			\addplot[black, style=dashed, line width=0.75pt,smooth,every mark/.append style={solid}, mark=otimes*, mark repeat=1] table[x index=0, y index=2]{HardLimLoss_ML_ISI3_m21dB.txt};
			\addlegendentry{$\chi^{\star} (-21\text{ dB})$};
			
			\addplot[red, style=solid, line width=0.75pt,smooth,every mark/.append style={solid}, mark=diamond*, mark repeat=1] table[x index=0, y index=1]{HardLimLoss_ML_ISI3_m6dB.txt};
			\addlegendentry{$\chi\phantom{^{\star}} (-6\text{ dB})$};
			
			\addplot[red, style=dashed, line width=0.75pt,smooth,every mark/.append style={solid}, mark=diamond*, mark repeat=1] table[x index=0, y index=2]{HardLimLoss_ML_ISI3_m6dB.txt};
			\addlegendentry{$\chi^{\star} (-6\text{ dB})$};
			
			\addplot[blue, style=solid, line width=0.75pt,smooth,every mark/.append style={solid}, mark=square*, mark repeat=1] table[x index=0, y index=1]{HardLimLoss_ML_ISI3_m3dB.txt};
			\addlegendentry{$\chi\phantom{^{\star}} (-3\text{ dB})$};

			\addplot[blue, style=dashed, line width=0.75pt,smooth,every mark/.append style={solid}, mark=square*, mark repeat=1] table[x index=0, y index=2]{HardLimLoss_ML_ISI3_m3dB.txt};
			\addlegendentry{$\chi^{\star} (-3\text{ dB})$};

	\end{axis}

\end{tikzpicture}
\caption{Quantization Loss - Deterministic ISI Channel}
\label{HardLimLoss_ISI}
\end{figure}
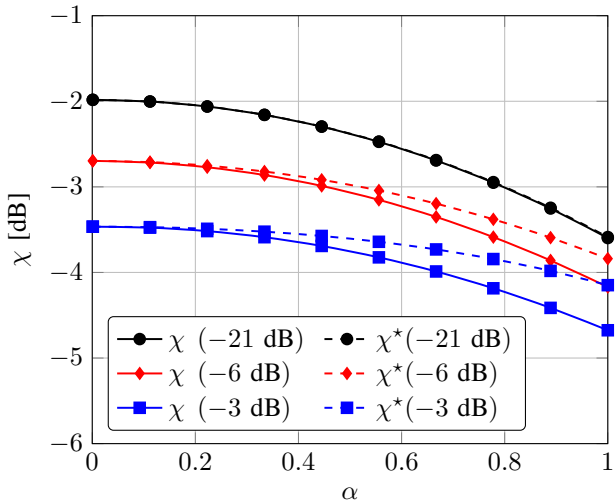
In Fig. \ref{HardLimLoss_ISI} we visualize the performance loss defined in \eqref{def:quantization:loss:det} and \eqref{def:quantization:loss:det:genius} due to hard-limiting the receive signal \eqref{def:ISI:quantized:receiver}. It can be observed that the loss is less pronounced in the low SNR setup while, in general, it increases with the quantization offset $\alpha$.
\begin{figure}[!htbp]
\begin{tikzpicture}[scale=1.0]

  	\begin{axis}[ylabel=$\Upsilon\text{ [dB]}$,
  			xlabel=$\alpha$,
			grid,
			tick label style={/pgf/number format/fixed},
			ymin=-0.6,
			ymax=0.2,
			xmin=0,
			xmax=1,
			legend pos= south west]
			
    			\addplot[black, style=solid, line width=0.75pt,smooth,every mark/.append style={solid}, mark=otimes*, mark repeat=1] table[x index=0, y index=3]{HardLimLoss_ML_ISI3_m21dB.txt};
			\addlegendentry{$\text{SNR}=-21\text{ dB}$};
			
			\addplot[red, style=solid, line width=0.75pt,smooth,every mark/.append style={solid}, mark=diamond*, mark repeat=1] table[x index=0, y index=3]{HardLimLoss_ML_ISI3_m6dB.txt};
			\addlegendentry{$\text{SNR}=-6\text{ dB}$};
			
			\addplot[blue, style=solid, line width=0.75pt,smooth,every mark/.append style={solid}, mark=square*, mark repeat=1] table[x index=0, y index=3]{HardLimLoss_ML_ISI3_m3dB.txt};
			\addlegendentry{$\text{SNR}=-3\text{ dB}$};
		
	\end{axis}

\end{tikzpicture}
\caption{Offset Loss - Deterministic ISI Channel}
\label{OffsetLoss_ISI}
\end{figure}
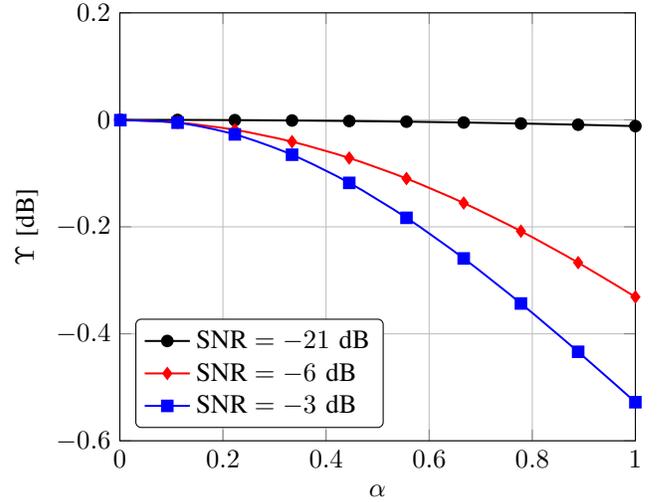
For the considered ISI scenario, the accuracy degradation due to the uncertainty in the unknown offset $\alpha$, shown in Fig. \ref{OffsetLoss_ISI}, is smaller than $-0.6$ dB for the considered range of offsets. In summary, the results show that for the wireless channel estimation task \eqref{def:ISI:quantized:receiver}, a quantization level $\alpha$ close to zero is, in general, preferable and that the performance gap between the ideal and the $1$-bit system increases with the SNR as well as with the offset value. Note, that in the low SNR regime, the fact that the offset is known to the receiver does not provide additional accuracy when estimating the ISI channel $\ve{\theta}$.
\subsubsection{Performance Analysis - Hybrid Approach}
For the case of a random channel, we assume $\boldsymbol{\theta}\sim\mathcal{N}(\mathbf{0}_K,\ve{R}_{\ve{\theta}})$, where $\mathbf{0}_K$ denotes the $K$-dimensional zero vector and $\ve{R}_{\ve{\theta}}\in\fieldR^{K \times K}$ is a diagonal matrix with $k$-th diagonal element $\sigma^2_{\theta_k}$.

With the ideal receiver, the asymptotic performance of the MAP estimator can be characterized by the ECRLB \eqref{eq:ECRLB}
\begin{align}\label{eq:ECRLB:ISI}
\text{MSE}_{\ve{y}}&\overset{a}{=}\exdi{\ve{\theta}}{\ve{F}^{-1}(\boldsymbol{\theta})}=\left(\sum_{n=1}^{N}  \ve{X}_n\right)^{-1}.
\end{align}
For the $1$-bit receiver, by plugging \eqref{FIM:theta:theta}-\eqref{FIM:alpha:alpha} into \eqref{eq:EHCRLB}, one obtains the EHCRLB. The quantization losses from \eqref{def:quantization:loss:hyp} and \eqref{def:quantization:loss:hyp:genius} are given by
\begin{align}
\chi(\alpha)&\overset{a}{=} \sum_{k=1}^{K}\frac{ \left[\left(\sum_{n=1}^{N}  \ve{X}_n\right)^{-1}\right]_{kk}}{\left\{\exdi{\ve{\theta}}{\bigg( \ve{J}_{\ve{\theta}\ve{\theta}}(\ve{\psi}) -  \frac{\ve{J}_{\ve{\theta}\alpha}(\ve{\psi})    \ve{J}_{\alpha\ve{\theta}}(\ve{\psi})}{{J}_{\alpha\alpha} (\ve{\psi})}   \bigg)^{-1}}\right\}_{kk}}\label{loss:rand:det:exp},\\
\chi^{\star}(\alpha)&\overset{a}{=} \sum_{k=1}^{K}\frac{\left[\left(\sum_{n=1}^{N}  \ve{X}_n\right)^{-1}\right]_{kk}}{\left\{\exdi{\ve{\theta}}{ \ve{J}^{-1}_{\ve{\theta}\ve{\theta}}(\ve{\psi})}\right\}_{kk}}\label{loss:rand:det:knownt}.
\end{align}
For low SNR, we identify that $\boldsymbol{\gamma}=\left[\begin{array}{ccc}
\sigma^2_{\theta_1} & \ldots & \sigma^2_{\theta_K}
\end{array}\right]$ and $\boldsymbol{\gamma}_0=\left[\begin{array}{ccc}
0 & \ldots & 0
\end{array}\right]$ to obtain the simplified expression
\begin{align}\label{eq:sim:hybrid:preMSE}
\lim_{{\ve{\gamma}\to\ve{\gamma}_0}}  \text{MSE}_{\ve{z}}(\alpha )&\overset{a}{=}\lim_{{\ve{\gamma}\to\ve{\gamma}_0}}  \text{MSE}^{\star}_{\ve{z}}(\alpha )\nonumber\\
&=\frac{1}{\phi_0(\alpha)}\left(\sum_{n=1}^{N}  \ve{X}_n\right)^{-1},
\end{align}
by using \eqref{def:ISI:fim:ideal:receiver} in \eqref{eq:SNR:hybrid:preMSE}.  
\subsubsection{Results - Hybrid Approach}
For the parameterization of the hybrid ISI channel with $K=3$, we use
\begin{align}
\sigma^2_{\theta_k}=\text{SNR}_k
\end{align}
and set the variances of the two interfering channel taps to $\text{SNR}_2=\text{SNR}_1-3\text{ dB}$ and $\text{SNR}_3=\text{SNR}_1-6\text{ dB}$.
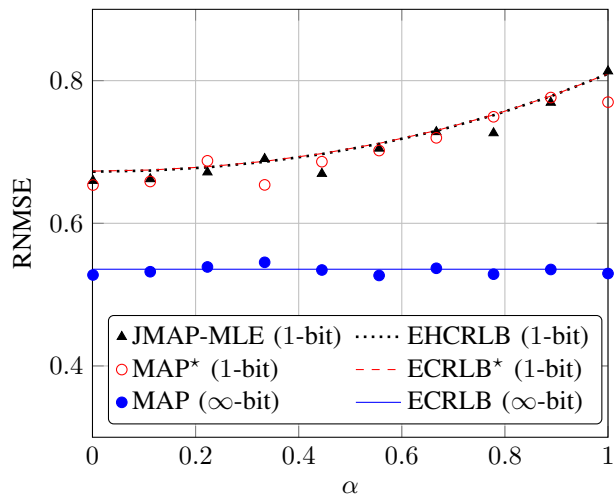
\begin{figure}[!htbp]
\begin{tikzpicture}[scale=1.0]

  	\begin{axis}[ylabel=$\text{RNMSE}$,
  			xlabel=$\alpha$,
			grid,
			tick label style={/pgf/number format/fixed},
			ymin=0.3,
			ymax=0.9,
			xmin=0,
			xmax=1,
			legend columns=2,
			legend style={/tikz/column 2/.style={column sep=3pt}},
			legend pos= south west]
						
			\addplot[black, only marks, mark=triangle*] table[x index=0, y index=1]{Algorithm_JMAPML_ISI3_1bit_m21dB.txt};
			\addlegendentry{JMAP-MLE ($1$-bit)};
			
			\addplot[black, style= dotted,line width=1pt] table[x index=0, y index=2]{Algorithm_JMAPML_ISI3_1bit_m21dB.txt};
			\addlegendentry{EHCRLB ($1$-bit)};
			
			\addplot[red, only marks, mark=o] table[x index=0, y index=1]{Algorithm_MAP_ISI3_1bit_genius_m21dB.txt};
			\addlegendentry{MAP$^{\star}$ ($1$-bit)};
			
			\addplot[red, style=dashed] table[x index=0, y index=2]{Algorithm_MAP_ISI3_1bit_genius_m21dB.txt};
			\addlegendentry{ECRLB$^{\star}$ ($1$-bit)};
			
			\addplot[blue, only marks, mark=*] table[x index=0, y index=1]{Algorithm_MAP_ISI3_ideal_m21dB.txt};
			\addlegendentry{MAP ($\infty$-bit)};
			
			\addplot[blue] table[x index=0, y index=2]{Algorithm_MAP_ISI3_ideal_m21dB.txt};
			\addlegendentry{ECRLB ($\infty$-bit)};
			
	\end{axis}

\end{tikzpicture}
\caption{MSE - Hybrid ISI Channel ($\text{SNR}=-21$ dB)}
\label{fig:PerformanceMAP:ISI:lowSNR}
\end{figure}
Fig. \ref{fig:PerformanceMAP:ISI:lowSNR} shows the performance of the quantized receiver in a low SNR scenario ($\text{SNR}=-21$ dB) with and without knowledge of the hard-limiting offset from \eqref{eq:MAP:quant:genius} and \eqref{eq:MAP:quant}, respectively. As a reference, the performance of the ideal receive system \eqref{eq:MAPy} is also plotted.
\begin{figure}[!htbp]
\begin{tikzpicture}[scale=1.0]

  	\begin{axis}[ylabel=$\text{RNMSE}$,
  			xlabel=$\alpha$,
			grid,
			tick label style={/pgf/number format/fixed},
			ymin=0.03,
			ymax=0.13,
			xmin=0,
			xmax=1,
			legend columns=2,
			legend style={/tikz/column 2/.style={column sep=5pt}},
			legend pos= south west]
			
			\addplot[black, only marks, mark=triangle*] table[x index=0, y index=1]{Algorithm_JMAPML_ISI3_1bit_m3dB.txt};
			\addlegendentry{MLE ($1$-bit)};
			
			\addplot[black, style=dotted, line width=1pt] table[x index=0, y index=2]{Algorithm_JMAPML_ISI3_1bit_m3dB.txt};
			\addlegendentry{EHCRLB ($1$-bit)};
			
			\addplot[red, only marks, mark=o] table[x index=0, y index=1]{Algorithm_MAP_ISI3_1bit_genius_m3dB.txt};
			\addlegendentry{MAP$^{\star}$ ($1$-bit)};
			
			\addplot[red, style=dashed] table[x index=0, y index=2]{Algorithm_MAP_ISI3_1bit_genius_m3dB.txt};
			\addlegendentry{ECRLB$^{\star}$ ($1$-bit)};
			
			\addplot[blue, only marks, mark=*] table[x index=0, y index=1]{Algorithm_MAP_ISI3_ideal_m3dB.txt};
			\addlegendentry{MAP ($\infty$-bit)};
			
			\addplot[blue] table[x index=0, y index=2]{Algorithm_MAP_ISI3_ideal_m3dB.txt};
			\addlegendentry{ECRLB ($\infty$-bit)};
			
	\end{axis}

\end{tikzpicture}
\caption{MSE - Hybrid ISI Channel ($\text{SNR}=-3$ dB)}
\label{fig:PerformanceMAP:ISI:mediumSNR}
\end{figure}
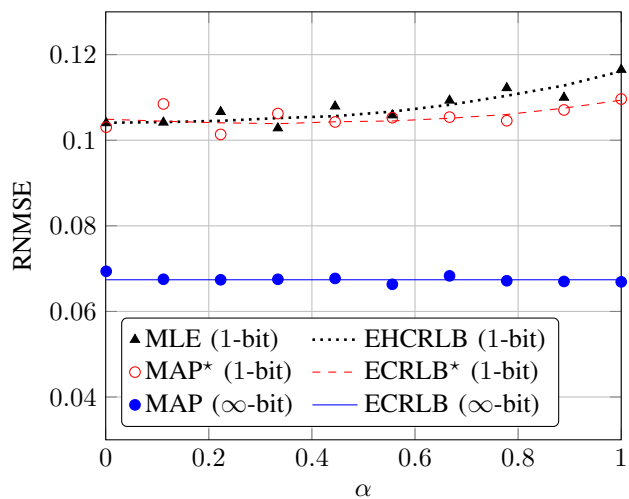
In Fig. \ref{fig:PerformanceMAP:ISI:mediumSNR} the RNMSE is depicted for a medium SNR setup ($\text{SNR}=-3$ dB). It can be observed that the analytic error formulas provide an accurate assessment of the behavior of the estimation algorithms in the ISI channel model. In Fig. \ref{HardLimLoss_ISI_MAP} we explicitly sketch the quantization loss, while in Fig. \ref{OffsetLoss_MAP_ISI} the accuracy degradation due to the estimation of the unknown offset is depicted. It can be observed that like in the deterministic case (Figs. \ref{HardLimLoss_ISI} and \ref{OffsetLoss_ISI}), the loss due to the unknown threshold is small for the considered range of offsets.
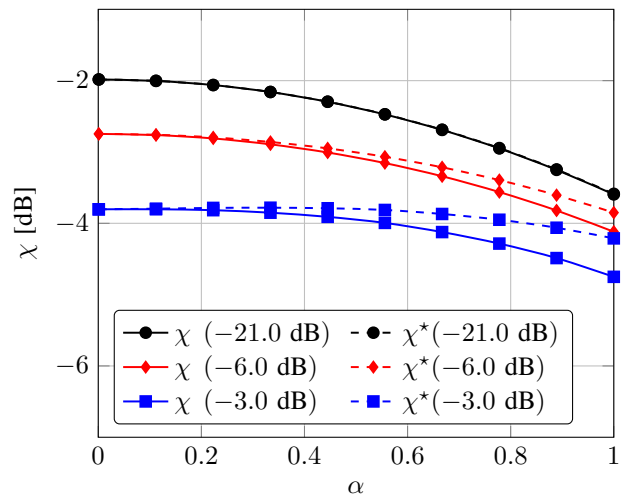
\begin{figure}[!htbp]
\begin{tikzpicture}[scale=1.0]

  	\begin{axis}[ylabel=$\chi \text{ [dB]}$,
  			xlabel=$\alpha$,
			grid,
			tick label style={/pgf/number format/fixed},
			ymin=-7.0,
			ymax=-1.0,
			xmin=0,
			xmax=1,
			legend columns=2,
			legend style={/tikz/column 2/.style={column sep=5pt}},
			legend pos= south west]
			
    			\addplot[black, style=solid, line width=0.75pt,smooth,every mark/.append style={solid}, mark=otimes*, mark repeat=1] table[x index=0, y index=1]{HardLimLoss_MAP_ISI3_m21dB.txt};
			\addlegendentry{$\chi\phantom{^{\star}} (-21.0\text{ dB})$};
			
			\addplot[black, style=dashed, line width=0.75pt,smooth,every mark/.append style={solid}, mark=otimes*, mark repeat=1] table[x index=0, y index=2]{HardLimLoss_MAP_ISI3_m21dB.txt};
			\addlegendentry{$\chi^{\star} (-21.0\text{ dB})$};
			
			\addplot[red, style=solid, line width=0.75pt,smooth,every mark/.append style={solid}, mark=diamond*, mark repeat=1] table[x index=0, y index=1]{HardLimLoss_MAP_ISI3_m6dB.txt};
			\addlegendentry{$\chi\phantom{^{\star}} (-6.0\text{ dB})$};
			
			\addplot[red, style=dashed, line width=0.75pt,smooth,every mark/.append style={solid}, mark=diamond*, mark repeat=1] table[x index=0, y index=2]{HardLimLoss_MAP_ISI3_m6dB.txt};
			\addlegendentry{$\chi^{\star} (-6.0\text{ dB})$};
			
			\addplot[blue, style=solid, line width=0.75pt,smooth,every mark/.append style={solid}, mark=square*, mark repeat=1] table[x index=0, y index=1]{HardLimLoss_MAP_ISI3_m3dB.txt};
			\addlegendentry{$\chi\phantom{^{\star}} (-3.0\text{ dB})$};

			\addplot[blue, style=dashed, line width=0.75pt,smooth,every mark/.append style={solid}, mark=square*, mark repeat=1] table[x index=0, y index=2]{HardLimLoss_MAP_ISI3_m3dB.txt};
			\addlegendentry{$\chi^{\star} (-3.0\text{ dB})$};

	\end{axis}

\end{tikzpicture}
\caption{Quantization Loss - Hybrid ISI Channel}
\label{HardLimLoss_ISI_MAP}
\end{figure}
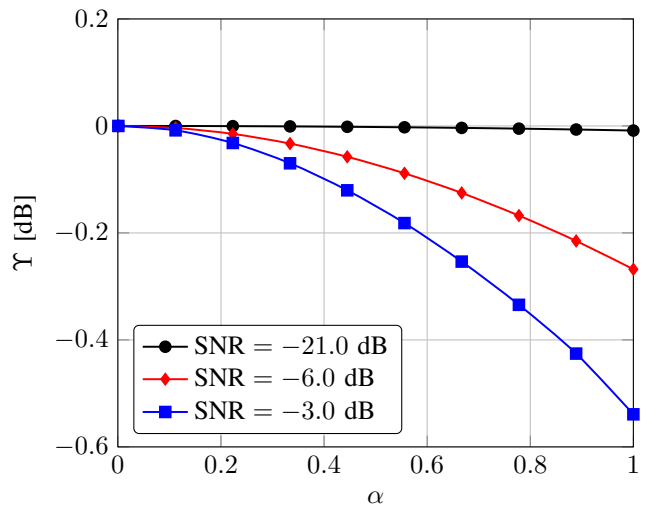
\begin{figure}[!htbp]
\begin{tikzpicture}[scale=1.0]

  	\begin{axis}[ylabel=$\Upsilon\text{ [dB]}$,
  			xlabel=$\alpha$,
			grid,
			tick label style={/pgf/number format/fixed},
			ymin=-0.6,
			ymax=0.2,
			xmin=0,
			xmax=1,
			legend pos= south west]
			
    			\addplot[black, style=solid, line width=0.75pt,smooth,every mark/.append style={solid}, mark=otimes*, mark repeat=1] table[x index=0, y index=3]{HardLimLoss_MAP_ISI3_m21dB.txt};
			\addlegendentry{$\text{SNR}=-21.0\text{ dB}$};
			
			\addplot[red, style=solid, line width=0.75pt,smooth,every mark/.append style={solid}, mark=diamond*, mark repeat=1] table[x index=0, y index=3]{HardLimLoss_MAP_ISI3_m6dB.txt};
			\addlegendentry{$\text{SNR}=-6.0\text{ dB}$};
			
			\addplot[blue, style=solid, line width=0.75pt,smooth,every mark/.append style={solid}, mark=square*, mark repeat=1] table[x index=0, y index=3]{HardLimLoss_MAP_ISI3_m3dB.txt};
			\addlegendentry{$\text{SNR}=-3.0\text{ dB}$};
		
	\end{axis}

\end{tikzpicture}
\caption{Offset Loss - Hybrid ISI Channel}
\label{OffsetLoss_MAP_ISI}
\end{figure}
\subsection{Single-tap SISO Channel Estimation}
For the special case of a single channel tap, i.e., $K=1$, the derived expressions can be further simplified \cite{SteinBar_ICASSP2016}. 
\subsubsection{Performance - Deterministic Approach}
For the deterministic case, with \eqref{def:ISI:fim:ideal:receiver} we obtain
\begin{align}
{F}({\theta})=N.
\end{align}
For the hard-limited receiver in \eqref{def:ISI:quantized:receiver}, we have
\begin{align}
\phi_n(\theta,\alpha)=\frac{ \expb{ -\big(\alpha- \theta x_n\big)^2 }  }{ {2\pi} \big( \qfunc{\alpha- \theta x_n} - \qfuncptwo{\alpha- \theta x_n} \big) }.
\end{align}
Therefore, using \eqref{FIM:theta:theta}-\eqref{FIM:alpha:alpha} one obtains
\begin{align}
J_{{\theta}{\theta}}(\theta,\alpha) &=\sum_{n=1}^{N} \phi_{n}(\theta,\alpha) {x}_n^2\notag\\
&=\frac{N}{2}\big(\phi_{+}(\theta, \alpha)+\phi_{-}(\theta, \alpha)\big),\label{info:fisher:1bit:siso:theta:theta}\\
J_{{\theta}\alpha}(\theta,\alpha)&=-\sum_{n=1}^{N} \phi_{n}(\theta,\alpha) {x}_n\notag\\
&= -\frac{N}{2}\big(\phi_{+}(\theta, \alpha)-\phi_{-}(\theta, \alpha)\big),\label{info:fisher:1bit:siso:theta:alpha}
\end{align}
and
\begin{align}
J_{\alpha\alpha}(\alpha\alpha)&=\sum_{n=1}^{N} \phi_{n}(\theta,\alpha)\notag\\
&=\frac{N}{2}\big(\phi_{+}(\theta, \alpha)+\phi_{-}(\theta, \alpha)\big),\label{info:fisher:1bit:siso:alpha:alpha}
\end{align}
where for brevity we define
\begin{align}
\phi_{+}(\theta, \alpha) &\triangleq \frac{ \expb{ -(\alpha + \theta)^2 } }{ {2\pi} \big(\qfunc{\alpha + \theta}-\qfuncptwo{\alpha + \theta}\big)},\\
\phi_{-}(\theta, \alpha) &\triangleq \frac{ \expb{ -(\alpha - \theta)^2 } }{ {2\pi} \big(\qfunc{\alpha - \theta}-\qfuncptwo{\alpha- \theta}\big)}.
\end{align}
Note, that the expressions \eqref{info:fisher:1bit:siso:theta:theta}-\eqref{info:fisher:1bit:siso:alpha:alpha} are due to the fact that with an equal symbol assignment each of the two BPSK signals is present for $\frac{N}{2}$ of the pilot symbols. Calculating the MSEs with \eqref{eq:CRLB}, \eqref{eq:CRLB:1bit}, and \eqref{eq:CRLB:1bit:genius}, we obtain
\begin{align}\label{eq:mse:SISO:unquantized:det}
\text{MSE}_{\ve{y}}(\theta)&\overset{a}{=}F^{-1}(\theta)=N^{-1}
\end{align}
and 
\begin{align}\label{eq:mse:SISO:1bit:det}
\text{MSE}_{\ve{z}}(\theta,\alpha)&\overset{a}{=}\bigg({J}_{\theta\theta}({\theta},{\alpha})-\frac{ {J}_{\theta\alpha}^2({\theta}, {\alpha}) }{{J}_{\alpha\alpha}({\theta},{\alpha})}\bigg)^{-1}\notag\\
&=\frac{1}{2N} \frac{ \phi_{+}(\theta, \alpha)+\phi_{-}(\theta, \alpha)}{ \phi_{+}(\theta, \alpha)\phi_{-}(\theta, \alpha)},\\
\label{eq:mse:SISO:1bit:genius:det}
\text{MSE}_{\ve{z}}^{\star}(\theta,\alpha)&\overset{a}{=}{J}_{\theta\theta}^{-1}({\theta},{\alpha})\notag\\
&=\frac{1}{2N} \big( \phi_{+}(\theta, \alpha)+\phi_{-}(\theta, \alpha) \big)^{-1}.
\end{align}

When comparing both receivers corresponding to the data models in \eqref{def:ISI:ideal:receiver} and \eqref{def:ISI:quantized:receiver}, the loss \eqref{def:quantization:loss:det} is given by
\begin{align}
\label{eq:loss:det}
\chi(\theta,\alpha)&\overset{a}{=}\frac{{J}_{\theta\theta}({\theta},{\alpha})}{{F}({\theta})}-\frac{ {J}_{\theta\alpha}^2({\theta}, {\alpha}) }{{J}_{\alpha\alpha}({\theta},{\alpha}){F}({\theta})}\notag\\
&=2 \frac{ \phi_{+}(\theta, \alpha)\phi_{-}(\theta, \alpha)}{ \phi_{+}(\theta, \alpha)+\phi_{-}(\theta, \alpha)}.
\end{align}
Assuming that the offset is known in \eqref{def:ISI:quantized:receiver} and using \eqref{def:quantization:loss:det:genius} we obtain
\begin{align}\label{eq:offset:loss:SISO:det}
\chi^{\star}(\theta,\alpha)&\overset{a}{=}\frac{{J}_{\theta\theta}({\theta},{\alpha})}{F(\theta)}\notag\\
&=\frac{1}{2}\big(\phi_{+}(\theta, \alpha)+\phi_{-}(\theta, \alpha)\big),
\end{align}
which, as predicted in \eqref{def:quantization:loss:det:genius:low:snr}, in the low SNR regime becomes
\begin{align}
\lim_{\theta\to0} \chi^{\star}(\theta,\alpha)&=\phi_0(\alpha).
\end{align}
As for a single channel parameter, i.e., $K=1$, with \eqref{def:loss:offset:hyp}
\begin{align}
\Upsilon(\theta,\alpha){=}\frac{\text{MSE}_{\ve{z}}^{\star}(\theta,\alpha)}{\text{MSE}_{\ve{z}}(\theta,\alpha)},
\end{align}
the asymptotic loss due to the uncertainty in the hard-limiter offset $\alpha$ in the data model \eqref{def:ISI:quantized:receiver} is
\begin{align}\label{eq:offset:loss:SISO:1bit:det}
\Upsilon(\theta,\alpha)&\overset{a}{=} \frac{4\phi_{+}(\theta, \alpha)\phi_{-}(\theta, \alpha)}{ \big( \phi_{+}(\theta, \alpha)+\phi_{-}(\theta, \alpha) \big)^2}.
\end{align}
As $\phi_{+}(0, \alpha)=\phi_{-}(0, \alpha)=\phi_{0}(\alpha)$, according to \eqref{offset:estimation:loss:lowSNR}, the ratio \eqref{eq:offset:loss:SISO:1bit:det} approaches $1$ in low SNR scenarios.

\begin{figure}[!htbp]
\begin{tikzpicture}[scale=1.0]

  	\begin{axis}[ylabel=$\Upsilon\text{ [dB]}$,
  			xlabel=$\alpha$,
			grid,
			ymin=-1.5,
			ymax=0.5,
			xmin=0,
			xmax=1,
			legend pos= south west]
			
    			\addplot[black, style=solid, line width=0.75pt,smooth,every mark/.append style={solid}, mark=otimes*, mark repeat=2] table[x index=0, y index=1]{HarLimLoss_DD_add.txt};
			\addlegendentry{$\Upsilon$ ($-21\text{ dB}$)};
			
			\addplot[red, style=solid, line width=0.75pt,smooth,every mark/.append style={solid}, mark=diamond*, mark repeat=2] table[x index=0, y index=2]{HarLimLoss_DD_add.txt};
			\addlegendentry{$\Upsilon$ ($-6\text{ dB}$)};
			
			\addplot[blue, style=solid, line width=0.75pt,smooth,every mark/.append style={solid}, mark=square*, mark repeat=2] table[x index=0, y index=3]{HarLimLoss_DD_add.txt};
			\addlegendentry{$\Upsilon$ ($-3\text{ dB}$)};
		
	\end{axis}

\end{tikzpicture}
\caption{Offset Loss - Deterministic Single-Tap Channel}
\label{HardLimLoss_DD_add}
\end{figure}
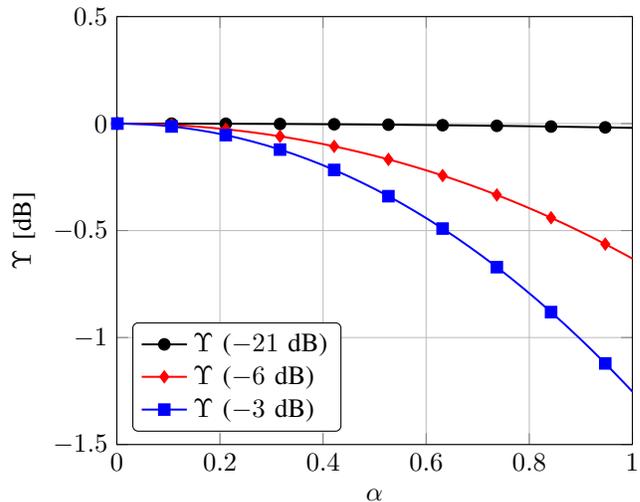

In Fig. \ref{HardLimLoss_DD_add} the performance loss in \eqref{eq:offset:loss:SISO:1bit:det} concerning the unknown offset $\alpha$ is visualized. While in the low SNR regime the estimation of $\alpha$ has almost no effect on the estimation of $\theta$, the situation changes within the medium SNR regime. Here the fact that the threshold is unknown can have a significant effect on to the estimation accuracy when the offset $\alpha$ is too far from the symmetric case. Interestingly, when comparing to the multi-tap loss in Fig. \ref{OffsetLoss_ISI}, it can be observed that the loss for the single-tap case is much more pronounced. This is because in the multi-tap channel the offset constitutes a significantly smaller portion of the parameter space $\ve{\Psi}$.
\subsubsection{Performance - Hybrid Approach}
In the case of a random channel parameter and $K=1$, the asymptotic performance of the MAP estimator with the ideal receiver, can be characterized using the ECRLB from \eqref{eq:ECRLB}
\begin{align}
\label{eq:ECRLB:SISO}
\text{MSE}_{\ve{y}} \overset{a}{=} \exdi{\theta}{F^{-1}(\theta)}=N^{-1}.
\end{align}
By plugging the expressions \eqref{info:fisher:1bit:siso:theta:theta}-\eqref{info:fisher:1bit:siso:alpha:alpha} into \eqref{eq:EHCRLB}, for the $1$-bit receiver, one obtains
\begin{align}
\label{eq:EHCRLB:1bit}
\text{MSE}_{\ve{z}}(\alpha)&\overset{a}{=}\exdi{\theta}{ \frac{1}{2N} \frac{ \phi_{+}(\theta, \alpha)+\phi_{-}(\theta, \alpha)}{ \phi_{+}(\theta, \alpha)\phi_{-}(\theta, \alpha)} }\notag\\
&= \frac{1}{2N}\Big(\exdi{\theta}{  \phi_{-}^{-1}(\theta, \alpha) }+\exdi{\theta}{  \phi_{+}^{-1}(\theta, \alpha) }\Big)\notag\\
&= \frac{1}{N} \exdi{\theta}{ \phi_{+}^{-1}(\theta, \alpha) },
\end{align}
where the last step holds due to symmetry, i.e.,
\begin{align}
\exdi{\theta}{ \phi_{-}^{-1}(\theta, \alpha) }=\exdi{\theta}{ \phi_{+}^{-1}(\theta, \alpha) }.
\end{align}
Under a known quantization threshold, we have 
\begin{align}
\label{eq:EHCRLB:1bit:genius}
\text{MSE}_{\ve{z}}^{\star}(\alpha)&\overset{a}{=}\frac{1}{2N}\exdi{\theta}{\big(\phi_{+}(\theta, \alpha)+\phi_{-}(\theta, \alpha)\big)^{-1}},
\end{align}
such that the asymptotic quantization losses are
\begin{align}
\chi(\alpha)&\overset{a}{=} \frac{1}{\exdi{\theta}{ \phi_{+}^{-1}(\theta, \alpha) }}\label{SSO:loss:rand:det:exp},\\
\chi^{\star}(\alpha)&\overset{a}{=}\frac{2}{\exdi{\theta}{ \big(\phi_{+}(\theta, \alpha)+\phi_{-}(\theta, \alpha) \big)^{-1}}}\label{SISO:loss:rand:det:knownt}.
\end{align}
For low SNR, we obtain the simplified expression
\begin{equation}
\lim_{\sigma_{\theta}^2\to0} \exdi{\theta}{  \phi_{+}^{-1}(\theta, \alpha) }= \phi_0^{-1}(\alpha),
\end{equation}
where the equality stems from the fact that the Gaussian density, parametrized by the continuous parameter $\sigma_{\theta}^2$ forms a positive summability kernel \cite[p. 9]{YK}.
Hence, the asymptotic performance loss in the low SNR domain is given by
\begin{equation}
\lim_{\sigma_{\theta}^2\to0} \chi(\alpha)\overset{a}{=}\phi_0(\alpha),
\end{equation}
such that the accuracy degradation due to the estimation of the unknown offset \eqref{def:loss:offset:hyp} is
\begin{align}\label{SISO:MAP:offset:loss}
\Upsilon(\alpha)&\overset{a}{=} \frac{\exdi{\theta}{ \big(\phi_{+}(\theta, \alpha)+\phi_{-}(\theta, \alpha) \big)^{-1}}}{2\exdi{\theta}{  \phi_{+}^{-1}(\theta, \alpha) }},
\end{align}
and vanishes in the low SNR regime, i.e.,
\begin{align}\label{SISO:MAP:offset:loss:vanishing}
\lim_{\sigma_{\theta}^2\to0} \Upsilon(\alpha)&\overset{a}{=}1.
\end{align}

\begin{figure}[!htbp]
\begin{tikzpicture}[scale=1.0]

  	\begin{axis}[ylabel=$\Upsilon\text{ [dB]}$,
  			xlabel=$\alpha$,
			grid,
			ymin=-1.5,
			ymax=0.5,
			xmin=0,
			xmax=1,
			legend pos= south west]
			
    			\addplot[black, style=solid, line width=0.75pt,smooth,every mark/.append style={solid}, mark=otimes*, mark repeat=1] table[x index=0, y index=3]{HardLimLoss_MAP_SISO_m21dB.txt};
			\addlegendentry{$\Upsilon$ ($-21\text{ dB}$)};
			
			\addplot[red, style=solid, line width=0.75pt,smooth,every mark/.append style={solid}, mark=diamond*, mark repeat=1] table[x index=0, y index=3]{HardLimLoss_MAP_SISO_m6dB.txt};
			\addlegendentry{$\Upsilon$ ($-6\text{ dB}$)};
			
			\addplot[blue, style=solid, line width=0.75pt,smooth,every mark/.append style={solid}, mark=square*, mark repeat=1] table[x index=0, y index=3]{HardLimLoss_MAP_SISO_m3dB.txt};
			\addlegendentry{$\Upsilon$ ($-3\text{ dB}$)};
		
	\end{axis}

\end{tikzpicture}
\caption{Offset Loss - Hybrid Single-Tap Channel}
\label{OffsetLoss_MAP_SISO}
\end{figure}
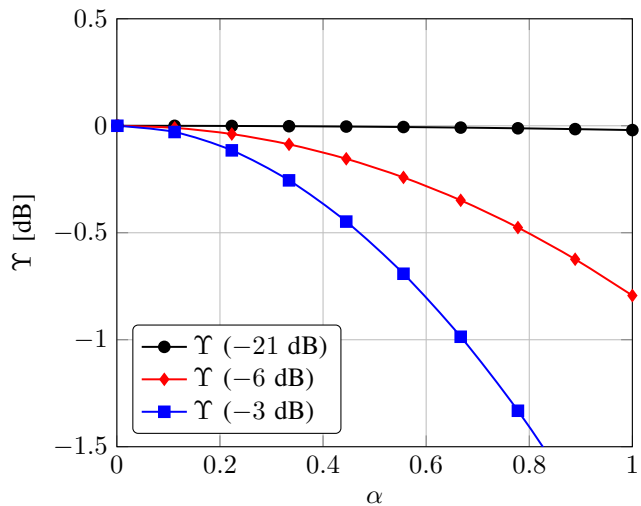

The accuracy degradation due to the unknown offset from \eqref{SISO:MAP:offset:loss} is visualized in Fig. \ref{OffsetLoss_MAP_SISO}. It shows that the offset estimation causes a significant additional error in the medium SNR regime while low SNR setups, as indicated by \eqref{SISO:MAP:offset:loss:vanishing}, the negative effect nearly vanishes. Also in the hybrid framework, it can be observed that  the single-tap offset loss (Fig. \ref{OffsetLoss_MAP_SISO}) is higher than in the multi-tap scenario (Fig. \ref{OffsetLoss_MAP_ISI}).
\subsection{Extension to Coarsely Quantized MIMO Channels}
The analysis in the paper can be extended to the case of MIMO channel, as follows. Under the assumption of spatially white sensor noise, one can consider a MIMO receiver being equivalent to $M_{R}$ independent multiple-input single-output (MISO) receive channels
\begin{align}\label{channel:model:mimo}
z_n^{(m)}  = \sign{ \ve{x}_n^{\T}\ve{\theta}^{(m)} + \eta_n^{(m)}  - \alpha^{(m)} },
\end{align}
where $\ve{x}_n$ denotes the pilot signals at the $M_{T}$ transmit antennas within the $n$-th symbol period, $\ve{\theta}^{(m)} $ the channel between the transmit antennas and the $m$-th receive antenna, $\eta_n^{(m)} $ the additive noise at the $m$-th receive antenna and $\alpha^{(m)} $ the corresponding $1$-bit ADC threshold. Due to the similarity between \eqref{channel:model:mimo} and the SISO multi-tap model \eqref{def:ISI:quantized:receiver} in Sec. \ref{sec:siso:multi:tap}, then also for the MIMO case the offset knowledge is not required in the low SNR regime (see Sec. \ref{sec:lowSNR}). Correspondingly, for medium SNR settings similar performance trends like for the multi-tap SISO channel (Sec. \ref{sec:siso:multi:tap}) are obtained for MIMO channels with unknown quantization thresholds.
\section{Conclusion}\label{sec:conc}
We have analyzed the problem of pilot-based channel parameter estimation from $1$-bit quantized data with an unknown hard-limiting threshold. In such a situation, in addition to the channel parameters, the receiver has to estimate the quantization level of the ADC. This has, in general, a negative impact on the achievable channel estimation accuracy. Providing a discussion for two different modeling approaches (deterministic and random channel parameters), we have shown analytically that, under mild conditions on the channel model and the pilot signal, lack of offset knowledge does in general not degrade the performance in the low SNR regime. Numerical results show that this conclusion also holds for medium SNR setups as long as the threshold of the $1$-bit quantizer is close to the symmetric case. For the ISI channel estimation problem with multiple channel taps, it was observed that the estimation loss due to an unknown offset is in general small while in the single-tap scenario the degradation is more pronounced. In summary, our findings confirm that $1$-bit A/D conversion is an attractive design option for future low-complexity wireless systems, in particular when the receiver is intended to solve complex channel estimation tasks in the low SNR regime. The presented results show that for such applications the requirements on the comparator circuit forming the low-complexity $1$-bit ADC are minor. Deviations of the offset from the symmetric case can be compensated at a small additional cost in the digital domain by appropriate estimation algorithms. For high-resolution signal processing with $1$-bit ADC in the medium SNR regime, our analysis shows that careful hardware design of the ADC is required, such that the comparator remains close to the symmetric case.
\appendices
\section{Derivation - FIM with $1$-bit ADC}\label{app:derivation:FIM}
Using the derivative \eqref{eq:condpdf:1bit:derivative} of the conditional probability mass function \eqref{eq:condpdf:1bit}, we obtain
\begin{align}\label{app:derivation:FIM:step1}
&\exdi{{z}_n|\ve{\psi}}{ \bigg(\frac{\partial \ln p_{{z}_n}({z}_n|\ve{\psi}) }{\partial \ve{\theta}} \bigg)^{\T} \frac{\partial \ln p_{{z}_n}({z}_n|\ve{\psi}) }{\partial \ve{\theta}} }=\notag\\
&=\exdi{{z}_n|\ve{\psi}}{ \frac{\exp{ \big(-(\alpha- s_{n}(\ve{\theta}))^2 \big)}  }{{2 \pi}\qfuncptwo{z_n(\alpha- s_{n}(\ve{\theta}))} }  \left( \frac{\partial s_{n}(\ve{\theta})}{\partial \ve{\theta}}\right)^{\T}  \frac{\partial s_{n}(\ve{\theta})}{\partial \ve{\theta}} }\notag\\
&=\frac{\exp{ \big(-(\alpha- s_{n}(\ve{\theta}))^2 \big)}  }{{2 \pi} } 
\exdi{{z}_n|\ve{\psi}}{ \frac{1}{\qfuncptwo{z_n(\alpha- s_{n}(\ve{\theta}))} } } \cdot\notag\\ 
&\cdot \left( \frac{\partial s_{n}(\ve{\theta})}{\partial \ve{\theta}}\right)^{\T}  \frac{\partial s_{n}(\ve{\theta})}{\partial \ve{\theta}}
\end{align}
Further, with \eqref{eq:condpdf:1bit} and the Q-function property $\qfunc{-\kappa}=1-\qfunc{\kappa},\kappa\in\fieldR$, the expectation in \eqref{app:derivation:FIM:step1} can be simplified
\begin{align}\label{app:derivation:FIM:step2}
&\exdi{{z}_n|\ve{\psi}}{ \frac{1}{\qfuncptwo{z_n(\alpha- s_{n}(\ve{\theta}))} }}=\notag\\
&= \frac{\qfunc{\alpha- s_{n}(\ve{\theta})}}{ \qfuncptwo{\alpha- s_{n}(\ve{\theta})} } +\frac{\qfunc{-(\alpha- s_{n}(\ve{\theta}))}}{ \qfuncptwo{-(\alpha- s_{n}(\ve{\theta}))} }\notag\\
&=\frac{1}{ \qfunc{\alpha- s_{n}(\ve{\theta})} } +\frac{1}{ 1-\qfunc{\alpha- s_{n}(\ve{\theta})} }\notag\\
&=\frac{1}{ \qfunc{\alpha- s_{n}(\ve{\theta})}-\qfuncptwo{\alpha- s_{n}(\ve{\theta})}  }.
\end{align}
With definition \eqref{definition:phin}, \eqref{app:derivation:FIM:step1} and \eqref{app:derivation:FIM:step2} lead to the result \eqref{derivation:FIM:quantized}.
\section{Proof - Theorem 1 (EHCRLB)}\label{app:EHCRLB}
\begin{IEEEproof}
	Since the sequence of MLEs is asymptotically uniformly integrable, then \cite{Vaart}
	\begin{multline}
	\label{eq:ConditionalEHCRLB}
	\lim\limits_{N\to\infty}{\Bigg|\exdi{\ve{z}|\ve{\psi}}{\big(\ve{\hat{\theta}}_{\ve{z}}(\ve{z})-\ve{\theta}\big) \big(\ve{\hat{\theta}}_{\ve{z}}(\ve{z})-\ve{\theta}\big)^{\T}}}\\-\bigg( \ve{J}_{\ve{\theta}\ve{\theta}}(\ve{\psi}) -  \frac{\ve{J}_{\ve{\theta}\alpha}(\ve{\psi})    \ve{J}_{\alpha\ve{\theta}}(\ve{\psi})}{{J}_{\alpha\alpha} (\ve{\psi})}   \bigg)^{-1}\Bigg|=0.
	\end{multline}
	Consequently, the total law of expectation implies that
	\begin{align}
	\label{eq:EHCRLBPromo}
	&\lim\limits_{N\to\infty}\text{MSE}_{\ve{z}}(\alpha)=\notag\\
	&=\lim\limits_{N\to\infty}\exdi{\ve{\theta}}{\exdi{\ve{z}|\ve{\psi}}{\big(\ve{\hat{\theta}}_{\ve{z}}(\ve{z})-\ve{\theta}\big) \big(\ve{\hat{\theta}}_{\ve{z}}(\ve{z})-\ve{\theta}\big)^{\T}}}\nonumber
	\\
	&=\lim\limits_{N\to\infty}\exdi{\ve{\theta}}{\bigg( \ve{J}_{\ve{\theta}\ve{\theta}}(\ve{\psi}) -  \frac{\ve{J}_{\ve{\theta}\alpha}(\ve{\psi})    \ve{J}_{\alpha\ve{\theta}}(\ve{\psi})}{{J}_{\alpha\alpha} (\ve{\psi})}   \bigg)^{-1}}.
	\end{align}
\end{IEEEproof}
\end{document}